\documentclass[prb,amssymb,superscriptaddress,twocolumn]{revtex4}
\makeatletter
\def\l@subsubsection#1#2{}
\makeatother
\usepackage{tikz-cd}
\usepackage{aliascnt}
\usepackage[ps,matrix,arrow,curve]{xy}
\xyoption{dvips}
\usepackage{psfrag,graphicx}
\usepackage{dcolumn}
\usepackage{bm}
\usepackage{amsfonts,amssymb,amsmath}
\usepackage{amssymb}
\usepackage{xcolor}
\usepackage{tikz}
\usetikzlibrary{decorations.markings}
\usetikzlibrary{arrows}
\usepackage{epstopdf}
\usepackage{mathbbol}
\usepackage{verbatim}
\usepackage{tikz-cd}
\usepackage[section]{placeins}
\let\conjugatet\overline
\usepackage{natbib}

\usepackage{amsthm}

\theoremstyle{definition}

\newtheorem{<main str>}{<Main str name>}

\newcommand{\be}{\begin{equation}}
\newcommand{\ee}{\end{equation}}
\newcommand{\bq}{\begin{eqnarray}}
\newcommand{\eq}{\end{eqnarray}}

\newcommand{\no}{\nonumber\\}

\newcommand{\ket}[1]{\left |#1 \right\rangle}
\newcommand{\bra}[1]{\left \langle #1 \right |}
\newcommand{\braket}[2]{\left \langle #1 \right| \left.#2\right\rangle}

\newcommand{\trr}{\triangleright}

\newtheorem{theorem}{Theorem}[section]

\newcommand{\One}{{\bf 1}} 
\newcommand{\CC}{{\mathbb C}}  

\usetikzlibrary{positioning,shapes.misc}
\usetikzlibrary{calc}
\tikzset{middlearrow/.style={
		decoration={markings,
			mark= at position 0.5 with {\arrow{#1}} ,
		},
		postaction={decorate}
	}
}
\tikzset{
    partial ellipse/.style args={#1:#2:#3}{
        insert path={+ (#1:#3) arc (#1:#2:#3)}
    }
}
\tikzset{offmiddlearrow/.style={
		decoration={markings,
			mark= at position 0.25 with {\arrow{#1}} ,
		},
		postaction={decorate}
	}
}
\begin{document}
\title{Topological phases from higher gauge symmetry in 3+1D}
\newcommand{\leedsm}{School of Mathematics, University of
  Leeds, Leeds, LS2 9JT, United Kingdom}
\newcommand{\leedsp}{School of Physics and Astronomy, University of
  Leeds, Leeds, LS2 9JT, United Kingdom}
\newcommand{\epsrc}{This work was supported by EPSRC under grant EP/I038683/1}

\author{Alex Bullivant}
\affiliation{\leedsm}
\author{Marcos Cal\c{c}ada}
\affiliation{\leedsm}
\affiliation{Departamento de Matem\'atica e Estat\'{\i}stica,
  Universidade Estadual de Ponta Grossa, Ponta Grossa, PR,
  Brazil}
\author{Zolt\'an K\'ad\'ar}
\affiliation{\leedsm}
\author{Paul Martin}
\affiliation{\leedsm} 
\author{Jo\~ao Faria Martins}
\affiliation{\leedsm}


\begin{abstract}
 	We propose an exactly solvable Hamiltonian for topological phases in $3+1$ dimensions utilising ideas from higher lattice gauge theory, where the gauge symmetry is given by a finite 2-group. We explicitly show that the model is a Hamiltonian realisation of Yetter's homotopy 2-type topological quantum field theory whereby the groundstate projector of the model defined on the manifold $M^3$ is given by the partition function of the underlying topological quantum field theory for $M^3\times [0,1]$. We show that this result holds in any dimension and illustrate it by computing the ground state degeneracy for a selection of spatial manifolds and 2-groups. As an application we show that a subset of our model is dual to a class of Abelian Walker-Wang models describing $3+1$ dimensional topological insulators.
\end{abstract}
\maketitle
\tableofcontents    
\section{Introduction}

Topological phases of matter have received considerable interest 
recently due to 
their practical applications 
related to various quantum Hall effect phenomena \cite{wen1990topological,tsui1982two,kane2005quantum,Fradkin}; and 
for the realisation of topological quantum computation \cite{kitaev2003fault,PachosB}.

Topological phases of matter have an underlying effective (infra-red limit) description given by a Topological Quantum Field Theory (TQFT) 
\cite{atiyah1988topological,LevinWen,Fradkin,witten1989}. 
Such theories are independent of the metric structure of space time, so low-energy physical processes are insensitive to local perturbations. Amplitudes of these physical processes are global quantities, topological invariants of the configuration space. In a canonical approach, the Hamiltonian is a sum of mutually commuting constraints, so the groundstate space is their joint eigenspace \cite{kitaev2003fault,Stringnet,kogutgt}. 
The degeneracy of the groundstate and types, fusion and braiding of possibly exotic excitations fully characterize such a theory 
\cite{KitaevHoney,Wangtqc}. 
An important property of TQFTs is negative corrections to the 
experimentally observable
entanglement entropy (due to global constraints on the correlations)
\cite{KitaevPreskill,LevinWen,Turner,bullivant2016entropic}.

In 2+1D, all phases of quantum matter with topological order are described by Chern-Simons-Witten/$BF$ theories \cite{lopez1998fermionic}, (twisted) quantum double (QD) models \cite{kitaev2003fault,hu2013twisted} and Levin Wen string nets \cite{Stringnet}. 

In 3+1D there are few known examples of TQFTs. As such there is limited knowledge of the kinds of observables and quasi-excitations which could be expected to exist in 3+1D topological phases of matter. So far in 3+1D only the Dijkgraaf-Witten topological gauge theory \cite{dijkgraaf1990topological,wan2014twisted} which describe symmetry protected topological phases and the Crane-Yetter TQFT\cite{crane1993categorical,WalkerWang,Simon} which describe topological insulators have been studied in the physical literature, (see also the very recent construction of TQFT based on a {\em G crossed} braided fusion category \cite{cui2015gauging}). 
Both TQFTs give observables which depend on at most the fundamental group and the signature of the space-time. They both support quasi-excitations given by point-particles with charge like quantum numbers and fermionic/bosonic mutual statistics and loop excitations which carry both charge and flux like quantum numbers\cite{Simon,wan2014twisted}. 

In this article we will describe a Hamiltonian formalism for a third type of TQFT, the Yetter homotopy 2-type TQFT\cite{yetter1993tqft,porter1998topological}. Yetter's TQFT uses a {\em 2-group} (equivalently a {\em crossed module}) to define a TQFT utilising the ideas of topological higher lattice gauge theory. Unlike the previous theories, such a theory is also sensitive to the homotopy 2-type information of the space-time, e.g. the second homotopy group. This feature is likely to be necessary to find non-trivial (meaning neither bosonic nor fermionic) representations of the 
{\em loop braid} 
group\footnote{This group is the 3+1D analogue of the braid group, which governs particle statistics in 2+1D.}\cite{baez2007exotic, fenn1997braid, kadar2014local}, since the evolution of loop excitations are 
world-sheets.    


The aim of this work is to better understand candidate theories for topological phases of matter in 3+1D space-time. Many questions understood in the 2+1D case are still open 
in 3+1D.
For example, what is the relation between the ground-state degeneracy and the number of quasi-particle excitations?
In 2+1D it is known that there exist a 1-1 correspondence between the ground-state degeneracy on the torus and the number of irreducible quasi-particle excitations, 
but the analogue in 3+1D with the ground-state degeneracy on the 3-torus and the number of irreducible excitations is known not to hold\cite{moradi2015universal,wan2014twisted,jiang2014generalized}.
Another related question is what are the topological quantum numbers needed to classify the ground-states and quasi-excitations in 3+1D. Such quantum numbers are expected to be related to the generators of the mapping class group of the 3-torus $SL(3,\mathbb{Z})$ but a full proof is still lacking.
In this paper we will outline a model of topological phases of matter in 3+1D with a 2-group gauge symmetry.
(In an accompanying article we
explicitly verify
the mathematical consistency of such models,
and also further discuss
the topological observables and quantum numbers of the model.) 

Defining Hamiltonian formalisms for TQFTs in 2+1D and 3+1D has been an
effective strategy
for understanding the spectrum of observables and physical manifestations of topological phases of matter
\cite{Stringnet,WalkerWang,kitaev2003fault,hu2013twisted,wan2014twisted}.
This is the approach taken in this manuscript, where we
will present a Hamiltonian formalism for the Yetter TQFT in 3+1D.
The structure of the text is as follows. 
We introduce the basic components of Higher Lattice Gauge Theory (HLGT) in section \ref{hlgti},
including the definition of a crossed module (section \ref{cm}).
This serves as a framework for the remaining structure.
We will then outline the Hamiltonian model in section \ref{Hami}. 
Section~\ref{Yetterrelation}
explains why our model is a Hamiltonian formalism of the 4D Yetter TQFT.
Finally we
describe an inclusion of a class of our Hamiltonian models into the class of Walker-Wang models\cite{WalkerWang,Simon}
(which form a Hamiltonian presentation of the Crane-Yetter TQFT)
in section \ref{WalkerWang}.

\section{Higher lattice gauge theory\label{hlgti}}
In this section we will establish the lattice formulation of higher gauge theories. These are  more complicated
than ordinary lattice gauge theory.
Instead of a group (the gauge group) on lattice edges,
here we need two groups, the group of holonomies of ordinary 
and higher gauge fields.
The latter types of holonomies
sit on plaquettes and 
can be thought to arise from surface integral of a non-Abelian 2-connection
\cite{BaezSchreiber}.
Beside the two groups,
the physical edge/plaquette geometry induces 
two maps between them, which satisfy certain compatibility
conditions.
The collection of this data is called 
crossed module (crossed modules are actually equivalent to
2-groups \cite{BaezLauda}) and it replaces the notion of the gauge
group in ordinary gauge theories.
Just as the structure of the gauge group
ensures that gauge-invariant and measurable quantities are independent of the choices made when defining the holonomy and the way
holonomies are composed,
it is the structure encoded in a crossed module 
which takes care of the same independence in a higher gauge theory. 

For the proof of independence, some algebraic topology 
is needed.
The proof will be published in a companion paper \cite{BCKMM2}.
Here we lean instead on existing work \cite{pfeiffer2003higher},
and only sketch the internal consistency of the theory we derive our Hamiltonian model from.
\subsection{\label{cm}Crossed modules}

Let $G$ and $E$ be groups, $\partial:E\rightarrow G$ a group homomorphism and $\triangleright$ an action of $G$ on $E$ by automorphisms
(i.e., the maps $G\times E\to E, (g,e)\mapsto g\trr e$ are homomorphisms for both variables). If the Peiffer conditions
\begin{align}
&\partial(g\triangleright e)=g\partial(e)g^{-1}\qquad&\forall g\in G,\forall e\in E,\label{CM1}\\
&\partial(e)\triangleright f=efe^{-1}\qquad&\forall e,f\in E\label{CM2}.
\end{align}
are satisfied then the tuple $(G,E,\partial,\trr)$ is called a {\em crossed module}.

An example is $(G,G,{\bf id},\trr)$ with $g\trr h=ghg^{-1}$, the
double ${\cal D}G$ of the group $G$ \cite{}.
Another example is $(G,{\rm AUT}(G),{\rm ad},\trr)$,  where ${\rm
	AUT}(G)$ is the automorphism group of $G$. Here ${\rm ad}$ sends a
$g\in G$ to conjugation by $g$, and the action is simply by evaluation
of ${\rm AUT}(G)$.
If $V$ is a representation of a group $G$ then we
can build a crossed module $(G,V,\partial,\trr)$ where $V$ is a
group as a vector space,
$\partial(V)=\{\One_G\}$ and where $\trr$ is the given action of $G$ on $V$. 

Now, we need a lattice, encoding the physical space of the theory.

\subsection{Lattices and lattice paths}
Given a manifold 
$x$ we write $bd(x)$ for the boundary
and $(x)$ for $x \setminus bd(x)$.
Given a set $K$ of subsets of a set we write $|K|_u$ for the union.
Given a $d$-manifold $M$, 
a {\em lattice} $L$ for $M$ is a  
set of subsets $L^i$, for each $i=0,1,2,3$,
where $x \in L^i$ is a closed topological $i$-disk embedded in $M$,
satisfying the following requirements,
with  $M^i := |\cup_{j=0}^i L^j |_u$. 
\vspace{0.1cm}

\hspace{-0.4cm}For $i=1,2,3$ and for $x,y \in L^i$ we have 
$bd(x) \subset M^{i-1}$;
$(x) \cap y = \emptyset$;
$(x) \cap M^{i-1} = \emptyset$. 
\vspace{0.1cm}

\hspace{-0.4cm}Finally, either  $d\leq 3$ and $M^3=M$ or 
an extension of $L$ exists so that $M^d=M$, with all additional cells
in $L^4$ or above.
\begin{itemize}
	\item An element in $L^0$ is a point of $M$, called a {\em vertex}.
	\item An element in $L^1$ is called an edge or a {\em track}.
	\item An element in $L^2$ is called a face or a {\em plaquette}.
	\item An element in $L^3$ is called a {\em blob}. 
\end{itemize}
A lattice for $M$ is essentially the same as a regular CW-complex
decomposition for $M$
(a CW-complex is said to be regular if 
each attaching map is an embedding \cite{Hatcher}). 
Examples are triangulations and cubulations \cite{cooper1988triangulating}. 

To describe field configurations succinctly, 
we need to give extra structure to the lattice. Let us fix a total order on 
$L^0$ denoted $<$. 
We give reference orientation to each element of $L^1$ such that the source vertex is  smaller than the target vertex. 
(Note that the lattice does not contain 1-gons.)
For every element of $p\in L^2$ 
we distinguish the smallest vertex $v_0(p)$ and fix an orientation for $p$ according to which, 
for $p$ with $n>2$ boundary edges, $v_1(p)<v_{n-1}(p)$ for the two neighbours of $v_0(p)$. 
The default {\em target} $t_p$ of $p$ is the edge whose source is
$v_0(p)$, 
target is $v_{n-1}(p)$, the default {\em source} $s_p$ of $p$ is 
the path with consecutive boundary vertices 
$v_0(p),v_1(p),v_2(p),\dots,v_{n-1}(p)$. 
In figures, if the target path of a face $p$ is an edge then 
we indicate the target edge by a double arrow, thus the default
case is:
\begin{center}\includegraphics[width=2.5cm]{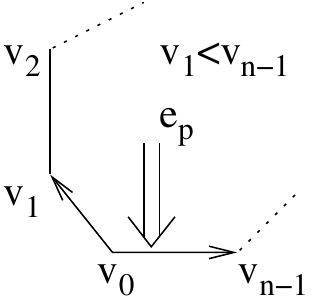}\end{center}
For simplicity here we exclude lattices with 2-gons in $L^2$.
Let us call the lattice with the chosen total order {\em dressed lattice}.

A simple path from vertex $v$ to vertex $v'$
in $L$ is a path in the 1-skeleton $M^1$
without repeated vertices. Thus a
simple path is a 1-disk in $M^1$ with its boundary decomposed into an
ordered pair of vertices (the source and target).
Similarly a 2-path is a disk surface $P$ in $M^2$ with $bd(P)$
decomposed into an ordered pair of simple paths. 

\newcommand{\CM}{(G,E,\partial, \trr)}  
\newcommand{\HL}{{\mathcal H}^L}  
\newcommand{\GG}{{\mathcal G}}   
\newcommand{\GGf}{\GG^L_{f\!\!f}}  
\newcommand{\ignore}[1]{}  

\subsection{Gauge fields}
Given a crossed module $(G,E,\partial, \trr)$ and a dressed lattice
$L$,
a gauge field configuration is an analogue of a conventional one,
which is encoded by 
a map $L^1 \rightarrow G$ assigning an element of gauge group
$G$ to each edge of $L$.
Beside 1-holonomies associated to directed paths, 
also 2-holonomies are associated to surfaces between two paths with common 
source and target.
Here it is encoded by functions $L^1\to G, i\mapsto g_i$ and 
$L^2\to E, p\mapsto e_p$. 
More precisely, we associate an element of $G$ to each oriented edge and an element of $E$ to each face  with
reference source and target. 
We call them the 1- and 2-holonomy with given source and target, respectively. 
Let us assume that a specific oriented edge $i$ is determined by its
vertices $v,w$ (with $v<w$).
Then we may write $i=vw$, and $g_i = g_{vw}$. 

We denote a complete `colouring' of the dressed lattice with such 
reference-oriented data by $L_c$;
and write $\GG^L = G^{|L^1|}\times E^{|L^2|} $ for the full set of colorings.

Given $i=vw \in L^1$, 
the 1-holonomy $g_{wv}$ with $w>v$ is given by $g_{wv} = g_{vw}^{-1}$. 
The 1-holonomy along a path in the 1-skeleton of the lattice is given
by the multiplication of the group elements of subpaths 
(and hence eventually of $g_i$s, 
or their inverses, depending on the direction of the edge with respect
to that of the path) along the path.
Note that 1-holonomy is well-defined by (existence of inverses and)
associativity of $G$. 
Similarly, we can 
compose 2-holonomies of disk-surfaces,
where the target of one coincides with the source of the next.
By (non-obvious) analogy with 1-holonomy, 2-path 2-holonomy is
well-defined by the crossed module axioms.
Below we explain the analogue of inverses and (more briefly) associativity.

First we will establish some notation. 
In figures, 2-holonomies of faces and other disk-surfaces will be
depicted by double arrows, which point toward the reference target
edge, just as for the 2-paths themselves.
For a face $p$ we 
will call the 1-holonomy $g_{s_p}$ of the source $s_p$ the source of
the 2-holonomy $e_p$. 
For example in the next figure we say that the 
2-holonomy associated to the triangle have source $g_1g_2$ and target $g_3$.
\begin{equation}
\raisebox{-0.5\height}{\includegraphics{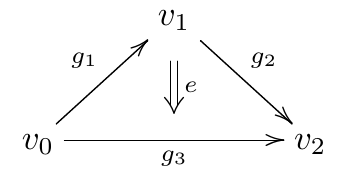}}
\end{equation}
Note, that our conventions for composing 1-holonomies throughout the
paper is $g_1g_2$ for consecutive edges $1$ and $2$.

In the following we will adopt a restriction on the lattice: we assume that there can be at most one edge between two vertices and that a face is determined by its boundary vertex set. This is not strictly necessary, but 
the notation becomes simpler, we will use $vw$ for the unique edge oriented from vertex $v$ to $w$ and $vwu$ for the unique 
triangle with distinct boundary vertices $v,w,u$.

\ignore{{ 
		In a face $p\in L^2$ with $n$ boundary edges the reference 2-holonomy
		is assumed to have target $g_{v_0\,v_{n-1}}$ and source
		$g_{v_0v_1}g_{v_1v_2}\dots g_{v_{n-2}v_{n-1}}$, 
		where $g_{vw}$ stands for $g_{wv}^{-1}$ wherever $w<v$. 
		Note, that these are the 1-holonomies
		of the default source and target of the face $p\in L^2$ of the dressed lattice.
		\begin{center}\includegraphics[width=2.5cm]{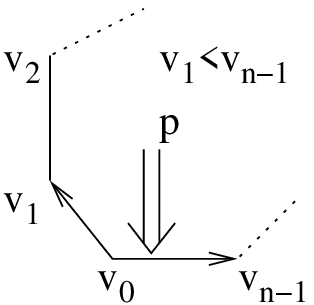}\end{center}
	}}

	%
	The terminology of source and target above comes from the axioms of
	2-categories, which is the language used in 
	Pfeiffer \cite{pfeiffer2003higher}, for example, 
	to define higher lattice gauge theories. In this paper,
	we will not define 2-categories, but simply write down  rules from that 
	formalism
	which we can use to define our model.
	
	We are now ready to compute the 2-holonomy of an arbitrary 2-path.
	As already noted, it is intrinsic to the notion of a gauge field that changing the direction of an edge is equivalent to changing 
	the associated group element to its inverse.
	One can compute the 1-holonomy along a path in the 1-skeleton of the lattice by using these transformations to ensure that 
	target of the 1-holonomy of an edge in the path coincides with the source of the next. 
	
	The 2-holonomy of a  
	2-path
	is constructed
	from the reference 2-holonomies of its plaquettes
	using a set of rules relating the reference 2-holonomy
	of each $p\in L^2$ with the 2-holonomy at $p$ with 
	different source and target.
	This way one
	multiplies all 2-holonomies of the elements of $L^2$ that are
	parts of the surface transformed appropriately so that the target of
	one is the source of the next. 
	
	The fact that the procedure is consistent and independent of
	the choices made will be explained in
	a companion paper \cite{BCKMM2} (using the language of crossed modules
	of groupoids equivalent to that of 2-groupoids). 
	Here we only illustrate the composition rules in indicative cases. 
	
	For each face $p\in L^2$ we define an 1-holonomy operator 
	\begin{equation}
	H_1(p)\equiv\partial e_p\,g_{s_p}\,g_{t_p}^{-1}\label{1hol}\end{equation}
	It is also called the {\em fake curvature} and it corresponds
	to the curvature 1-form\cite{GPP} of
	higher gauge theory\footnote{In the differential formulation
		of higher gauge theory, the equations of motion analogous to the vanishing
		of the field strength in ordinary gauge theory has an additive
		contribution of the derivative of the map $\partial$. So whenever the 
		latter is non-trivial the equation
		$H_1(p)=\One$ 
		is not equivalent to flatness of
		the 1-connection, hence the adjective "fake".
	}.
	In what follows, we will only consider configurations
	(unless otherwise stated) 
	where  $H_1(p)=\One\in G$ for $p\in L^2$.
	This is needed for consistency of
	the lattice formulation
	of 2d holonomy.
	
	Let us write down the multiplication convention and the
	rules of changing source and target of 2-holonomies. 
	We can multiply (compose) the 2-holonomy $e$ with $e'$ if $t_e=s_{e'}(=g_2$ in the figure):
	\begin{equation}
	\label{eq_vertical}
	\includegraphics{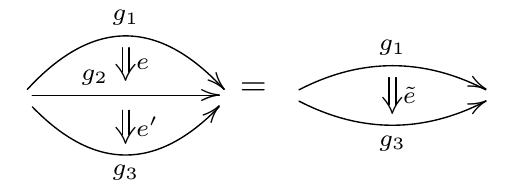}
	\end{equation}
        We use the convention to multiply the group elements from right to left $\tilde e=e^\prime\cdot e$.
	The 'whiskering' rules\cite{pfeiffer2003higher} for changing the source and target of a 2-holonomy are as follows:
	\begin{itemize}
		\item We can switch source and target of the 2-holonomy by changing $e$ to $e^{-1}$.
		\item We can change the direction of the source $g_s\in G$ and target $g_t\in G$ of the 2-holonomy simultaneously by changing 
		$e$ to $g_s^{-1}\trr e^{-1}$.
		\item We can change the source of both $s_e$ and $t_e$ 
		simultaneously and also the target of both as shown in the figure.
	\end{itemize}
	\begin{equation}
	\label{eq_whisker}\raisebox{-0.5\height}{\includegraphics{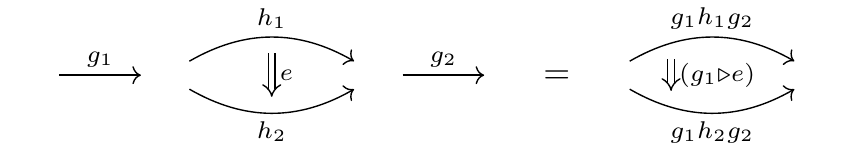}}
	\end{equation}
	Note, that $g$ ($g')$ can be the 1-holonomy of an edge or path in the boundary of the face, whose corresponding 2-holonomy 
	is $e$ as long as its target (source) is the source (target, respectively) of the 1-holonomies $s_e, t_e$. 
	(the source and target of $e$).   
	Here is an example of changing the source $s$ and target $t$ from the reference ones $s=g_1g_2, t=g_3$ of the 2-holonomy $e$ 
	associated to the triangle depicted in the first figure. 
	In the second figure $s=g_3g_2^{-1}, t=g_1$, in the third $s=g_1^{-1}g_3, t=g_2$. The transformation of the value of the 2-holonomy
	written on the double arrow is computed using the above described rules.
	\begin{widetext}
	\begin{equation}
	\raisebox{-0.5\height}{\includegraphics{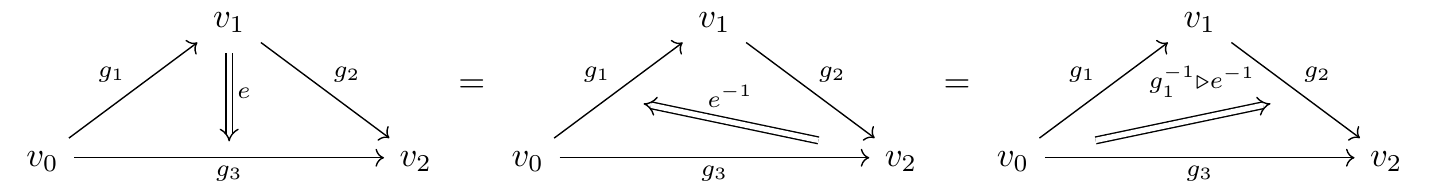}}
	\end{equation}
	\end{widetext}
	
	Note that the {\it fake flatness}  $H_1(p)={\bf 1}\in G$
	(the lhs. is defined by (\ref{1hol})) of the face $p$
	is a crucial condition for consistency of the above: changing the basepoint (the source of
	the source and target) of the 2-holonomy around the boundary of face $p$ back to the beginning gives
	\[g_ng_{n-1}^{-1}g_{n-2}^{-1}\dots g_1^{-1}\trr e_p=\partial e_p \trr e_p=e_pe_pe_p^{-1}=e_p\]
	by the second Peiffer condition of crossed modules (\ref{CM2}).
	The next example illustrates the composition of 2-holonomies associated to faces with a common boundary edge, where the condition
	of matching first target and second source is not satisfied.
	\begin{center}\includegraphics[width=2.5cm]{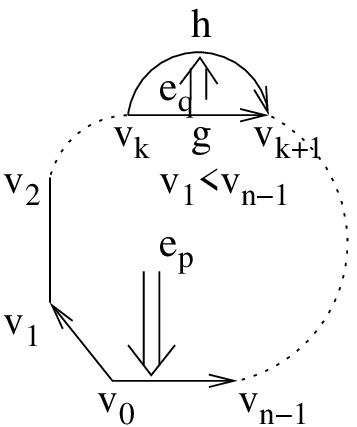}\end{center}
	Using the whiskering rules above, we can change the target and source
	of the top 2-holonomy $e_q$ to 
	$t(e'_q)=g_{v_0\,v_k}gg_{v_{k+1}\,v_{n-1}}$ and
	$s(e'_q)=g_{v_0\,v_k}hg_{v_k\,v_{n-1}}$, respectively, by changing $e_q$ to $e'_q\equiv g_{v_0\,v_k}\trr e_q^{-1}$, so the holonomy of the big disk 
	$r=p\cup q$ with $s(r)=g_{v_0\,v_k}hg_{v_k\,v_{n-1}}$ and $t(r)=g_{v_0 v_{n-1}}$ is
	\begin{equation} e_r=e_p(g_{v_0\,v_k}\trr e_q^{-1})\ .  \label{alon}\end{equation}
	Note, that $g_{vw}$ for a non-adjacent vertex pair $(v,w)$ stands for the 1-holonomy along the boundary of 
	the face according to its circular orientation. 
	
	We could use the whiskering rule to the direction opposite to the orientation of
	$p$, which results in 
	\begin{equation}
	(g_{\overline{v_0\,v_k}^o}\trr e_q^{-1})e_p\ ,\label{oppo}\end{equation}
	where $g_{\overline{vw}^o}$ denotes the 1-holonomy from $v$ to $w$
	along the boundary of $p$ in direction
	opposite to its circular orientation ($g_{\overline{v_0\,v_k}^o}=g_{v_0\,v_{n-1}}g_{v_{n-1}\,v_{n-2}}\dots g_{v_{k+1}v_k})$.
	Due to the fake flatness condition  
	the expressions (\ref{alon}) and (\ref{oppo}) agree:
	\begin{equation}\begin{array}{l}
	(g_{\overline{v_0\,v_k}^o}\trr e_q^{-1})e_p=(\partial e_p\,g_{v_0\,v_k}\trr e_q^{-1})\,e_p=\\\\
	e_p(g_{v_0\,v_k}\trr e_q^{-1}) e_p^{-1} e_p=e_p(g_{v_0\,v_k}\trr e_q^{-1})\ .\label{rotation}\end{array}\end{equation}
	In the first equality we used fake flatness $\partial e_p=g_{\overline{v_0\,v_k}^o}g_{\overline{v_k\,v_o}^o}$, in the second we used (\ref{CM2}). 
	
	Finally we give the formula for 
	the 2-holonomy operator 
	$H_2(P): \GG^L \rightarrow E$ in the case 
	of a reference tetrahedron $P \in L^2$:
	\[P\equiv \{[abcd],s(P)=t(P)=ad, a<b<c<d\}\]
	\begin{equation}\label{reftet}
	\includegraphics[width=\columnwidth]{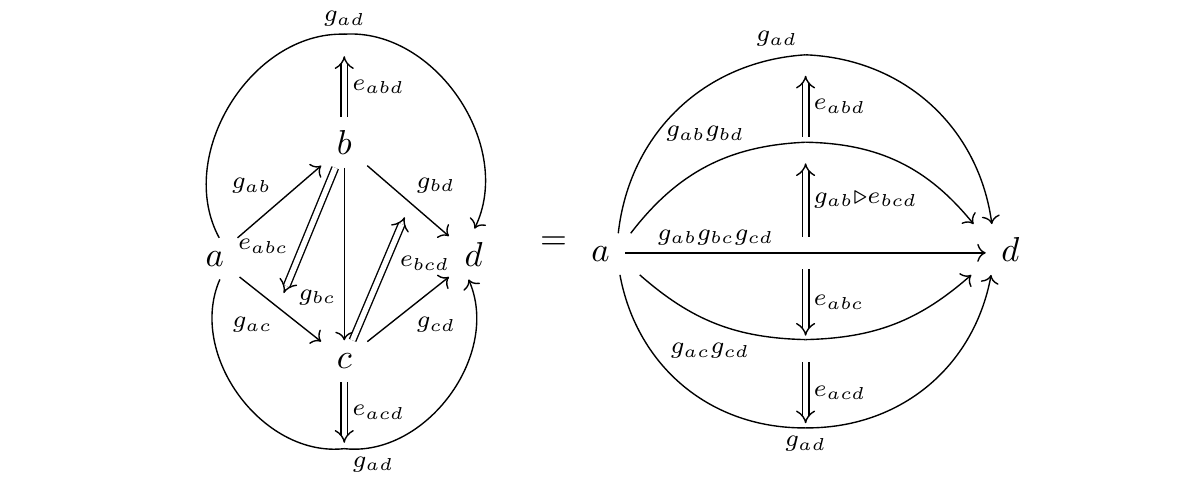}
	\end{equation}
	Using the multiplication convention we read the formula off from the figure
	\begin{equation}
	H_2(P)=e_{acd}e_{abc}(g_{ab}\triangleright e_{bcd}^{-1})e_{abd}^{-1} \ .\label{2hol}
	\end{equation}
	Note that choosing the basepoint to be the lowest ordered of
	the constituting vertices ($a$),
	and the direction of the 
	2-holonomy associated to the triangle $acd$
	to be that of the 2-holonomy associated to the tetrahedron,
	makes the latter unambigously defined.
	This fact follows from the work \cite{BCKMM2}
	(as discussed there, it  
	can be considered as a consequence of the Coherence Theorem for 2-categories).
\section{The Hamiltonian model \label{Hami}}
Recall that given a crossed module $\GG=\CM$ and a dressed lattice $L$
we have a set 
$ \GG^L$
of gauge field configurations or `colourings of $L$'.
We write ${\cal H}^L$
for the  `large' Hilbert space which has basis $\GG^L$ as a 
$\CC$-vector space, and  
has the natural delta-function scalar product.
We write
$  \left. \Big|\bigotimes_{i\in L^1} g_i\bigotimes_{p\in  L^2}e_p
\right\rangle
$
for  a colouring regarded as an element of $\HL$.

We also have the subset $\GGf$ of $\GG^L$  of fake flat colourings. 
The Hilbert space ${\cal H}$
is defined using $\GGf$ as basis.
These are the colourings satisfying the constraint at every face
$p \in L^2$ that
the fake curvature (\ref{1hol}) vanishes:
\[\begin{array}{l}
H_1(p)=\partial e_p\,g_{s_p} g_{t_p}^{-1}
=\\\\\partial e_p g_{v_0v_1}g_{v_1v_2}g_{v_2v_3}\dots g_{v_{n-2}v_{n-1}}
 g^{-1}_{v_{0}v_{n-1}}
= \One_{G}.\end{array}
\]

\ignore{
	The ``large" Hilbert space is
	${\cal H}^L={\mathbb C}(G^{|L^1|}\otimes E^{|L^2|})$,
	its basis corresponding to all
	possible colourings of the lattice $L^c$ reads
	\[
	\left\{ \left. \Big|\bigotimes_{i\in L^1} g_i\bigotimes_{p\in  L^2}e_p
	\right\rangle\right\}\ ,
	\]
	and the scalar product is the trivial one:
	\begin{equation}
	\left\langle \bigotimes_{i\in L^1} g_i\bigotimes_{p\in L^2} e_p\,\Big|\,\bigotimes_{i\in L^1} g'_i\bigotimes_{p\in L^2} e'_{p}
	\right\rangle=\prod_{i\in L^1}\delta_{g_i,g'_i}\prod_{p\in L^2}\delta_{e_p,e'_p}\ .
	\end{equation}
	The Hilbert space ${\cal H}$ that we will mostly consider is a subspace of the above, it is spanned by all colorings
	as in ${\cal H}^L$, but also subject to $|L^2|$ constraints of
	vanishing fake curvature at each face.
}
For example for a fake-flat colouring of $L$ a
`triangle' we may choose $e_p\in E$, $g_{v_0v_1},g_{v_1v_2}\in G$
arbitrarily, but then
$g_{v_0v_2}=\partial e_p\,g_{v_0v_1}g_{v_1v_2}$ is fixed.
Thus $\mbox{dim}({\cal H}) = |G|^2 |E|$ here.
The scalar product for ${\cal H}$ is the
one induced by that of ${\cal H}^L$.

In the following we will often use a simplified notation $|L_c\rangle$ for a basis
element of the Hilbert space, with $L$ denoting the dressed lattice
and $c$ its colouring. If clear from the context, 
we will also use this simplified notation for various dimensions and Hilbert spaces, for example for 
the QD models\cite{kitaev2003fault}, which corresponds to the
finite 'group' crossed module with 
$E=\{\bf 1\}$ and the restriction to $\CC (G^{|L_1|})$
forgetting about the trivial factor corresponding to face colouring.
There the lattice $L$
(or rather the underlying manifold)
is two dimensional.
\subsection{Gauge transformations}
We are now going to define operators in End$({\cal H}^L)$.
We will show in Section \ref{co2fl} that they restrict to End$({\cal H})$.
We will show how 
the 1- and 2-holonomy transform under their action,
and hence
show that they are gauge transformations.
We adopt the latter terminology now.

An intuitive way to think about these operators is
as follows.  
A {\it vertex} or 1-gauge {\it transformation}
at vertex $v$ is the analogue of ordinary $G$ gauge
transformation: edge labels change as they do in ordinary gauge
theory.
There, the 1-holonomies corresponding to boundaries of faces, also called Wilson loops, transform
by conjugation, their traces are observables.

In higher gauge theory however, the '1-holonomy' is already different: Compare $g_{ab}g_{bc}g_{ca}$ with $\partial(e_{abc})g_{ab}g_{bc}g_{ca}$.

Since the 2-holonomy does not ``really" (i.e. physically) change 
under 1-gauge transformations, the face labels are invariant except
when the vertex $v$ is the basepoint of the face.
An {\it edge transformation}
is a ``pure'' $E$ 2-gauge transformation:
it changes the 2-holonomy associated to each face $p$ adjacent to the edge. 
The action on the face 2-holonomy $e_p$ is the composition of an auxiliary face 2-holonomy also adjacent to the edge, 
where the source and target is appropriately modified to be composable with $e_p$. Here the quantities, which transform in a covariant way,
are the 2-holonomies associated to boundaries of blobs.
The edge label to which the edge  
transformation is 
associated also changes such that the auxiliary bigon composed of the edge and the transformed edge is fake flat.      

The transformation properties of 1-holonomies associated to faces, $H_1(p)$,
and 2-holonomies associated to blobs, $H_2(P)$, 
will be discussed in the next subsection.
There the reader can also find figures illustrating the effect of
the gauge transformations on a reference triangle. 
The explicit transformation formulas for the 1-holonomy of a triangle face
and the 2-holonomy of a tetrahedron  
are given in Appendix \ref{b}. 

Let us recall first the definition of left and right
multiplication operators for a group $G$
\begin{eqnarray*}
	L^g:&G\to G,&h\mapsto gh\\ R^g:&G\to G,&h\mapsto hg^{-1}
\end{eqnarray*}
linearly extended to the group algebra $\CC G$.
In the following we will use the notation $L_i^g$ ($R_i^g$), $i\in L^1, g\in G$
for the linear operator
in End$({\cal H}^L)$, which acts as left (right) multiplication by $g$
on the tensor factor $\CC G$ of ${\cal H}^L$ corresponding the
edge $i$ and identity on all other factors --- i.e. `locally' at $i$.
Similarly $L_p^e$ ($R_p^e$) stands for the same type of local
operators in 
End$({\cal H}^L)$ acting on the tensor factor $\CC E$
corresponding to the face $p$ and identity on all other factors. We
will also use $g\trr_p\,(\cdot)$ for the operator
acting on the tensor factor $E$ of ${\cal H}^L$ corresponding to
the face $p$ as
$e_p\mapsto g\trr e_p$.
The gauge transformation associated to vertex $v$ is defined by
\[ A_v^g=\prod_{i\in \star{(v)}}{\cal L}^g_v(i)\prod_{p\in \star{'(v)}}{\cal L}^g_v(p)\]
where $\star(v)$ ($\star{'(v)}$) is the set of reference edges (faces) adjacent to the vertex $v$, respectively;
the terms in the product are defined as follows:
%
\[
{\cal L}^g_v(i)=\left\{
\begin{array}{ll}
L_i^g, & \hbox{if $v=s(i)$;}\\
R^{g}_i, & \hbox{if $v=t(i)$.}
\end{array}
\right.
\]
\[
{\cal L}_v^g(p) = g \trr_p (\cdot), \,  \hbox{ if $v= v_0(p)$}
\]
and both families of operators act as identity in all other cases.
Note that all factors in the product of the expression of $A_v^g$ act on different tensor factors and their action depend
only on the parameter $g\in G$, so they commute: $A_v^g$ is well defined. The gauge transformation associated to edge $i$ is
defined by
\[{A}^e_i=L_i^{\partial e}\prod_{p\in\star{(i)}}{\cal L}^e_i(p)\]
where $\star(i)$ is the set of reference faces adjacent to the edge
$i$ and, for face $p$ an $n$-gon and 
$k, k+1 \in \{0, \cdots, n\}$,
\begin{equation*}
{\cal L}^e_i(p)= \left\{
\begin{array}{ll}
R^{g_{v_0v_k}\trr e}_p, & \hbox{if $i=v_k v_{k+1}$;} \\
L^{g_{\overline{v_0v_{k+1}}^o}\trr e}_p, & \hbox{if $i=v_{k+1}v_k$.}
\end{array}
\right.\ ,
\end{equation*}
where $g_{\overline{v_0v_l}^o}$ was defined at (\ref{rotation}).
Here $g_{v_0v_0} \equiv \One_G$,
$g_{v_0v_k}\equiv g_{v_0v_1}g_{v_1v_2}\dots g_{v_{k-1}v_k}$,
for $k=1, \cdots, n-1$,
$g_{v_0v_n}\equiv g_{v_0v_1}g_{v_1v_2}\dots g_{v_{n-1}v_0}$,
and we use the convention $i= v_nv_{n-1} = v_0v_{n-1}$.
Furthermore, all vertex labels are relative to the face $p$, but
the notation $v_k(p)$ instead of $v_k$ above would be too complicated.
Recall that each face $p$ ($n$-gon) of the lattice has a smallest vertex and a cyclic orientation $v_0,v_1,v_2,\dots,v_{n-1}$,
where $v_1$ is the smaller of the two neighbours of $v_0$ in $p$.
An earlier remark is also recalled here: if, in the labeling of the face $p$ for $0<l<n-2$ we have
$v_l>v_{l+1}$ in the previous sequence of group elements, then $g_{v_lv_{l+1}}=g_{v_{l+1}v_l}^{-1}$, with $g_{v_{l+1}v_l}$
being the reference 1-holonomy (often called colour in the following) of the corresponding edge.

As noted earlier, the crossed module with $E=\{\bf 1\}$ corresponds to the QD model\cite{kitaev2003fault} with group $G$, 
(but in dimension $d=3$ here). The set of gauge transformation are the vertex transformations defined above forgetting the 
trivial factor corresponding to the edge labels. 

The operators defined above satisfy the following relations for any vertices $v$,$v'$, any edges $i, i'$, any elements $g, h \in G$, and $e,f \in E$:
\begin{align}
&A^{g}_{v}A^{h}_{v}=A^{gh}_{v},    \label{gr1}\\
&A^{g}_{v}A^{h}_{v'}=A^{h}_{v'}A^{g}_{v},  & \hbox{if $v\neq v'$} \label{nb1}\\
&{ A}^{e}_i { A}^{f}_i={ A}^{ef}_{i},    \label{gr2}\\
&{ A}^{e}_i { A}^{f}_{i'}={ A}^{f}_{i'} { A}^{e}_i,  & \hbox{if $i \neq i'$} \label{nb2}\\
&A^g_v {A}^{e}_i = { A}^{e}_i A^g_v,   & \hbox{if $v \neq s(i)$} \label{mixe}\\
&A^g_v { A}^e_i = {A}^{g \trr e}_i A^g_v,  & \hbox{if $v=s(i)$.}  \label{mixh}
\end{align}
The proof of these identities are given in Appendix \ref{a}. 

\subsection{\label{co2fl}Covariance and 2-flatness constraints}
We will now define the operators enforcing `2-flatness' of boundaries of 
blobs 
$x\in L^3$. 

At this point we restrict the 
lattice to be a triangulation. The reason is of technical nature: for triangulations all blobs are tetrahedra and we have an essentially unique expression for their 2-holonomy given by (\ref{2hol}). Even for a cubic lattice, we would need to write a formula with  several distinct cases depending on the order of the vertices on the boundary of cubes. 
From a physical point of view, however, restricting ourselves to 
triangulations is not severe. 
(Nevertheless this restriction will be eliminated in our companion paper\cite{BCKMM2} where we will prove that any lattice can be used.)

\ignore{{
		For the 2-holonomy of a tetrahedron we can prove in a 
		relatively simple way that it 
		transforms covariantly under {\em gauge} transformations. We will do this also for the 1-holonomy. Hence the terminology of 
		`gauge transformation' is justified.
	}}
	
	The linear operators enforcing 
	2-flatness (trivial 2-holonomy) of a tetrahedron $P$($=bd(x)$ with
	$x\in L^3$); and fake-flatness, regarded as elements of End$({\cal H}^{L})$
	act on  basis elements in  $\mathcal{G}^{L}$ by
	\begin{equation}
	B_P=\delta_{H_2(P),\One}\ ; \qquad B_p=\delta_{H_1(p),\One}.
	\label{Bop}
	\end{equation}

	Let us now check how the holonomies 
	$H_1(p)=(\partial e_{abc})\,g_{ab}\,g_{bc}\,g_{ac}^{-1}$ and 
	$H_2(P)=e_{acd}\,e_{abc}\,(g\trr e_{bcd}^{-1})\,e_{abd}^{-1}$ transform under gauge 
	transformations. 
	The six gauge transformations
	`touching' 
	a triangle with vertices $a<b<c$ are depicted below. Caveat: Here and hereafter, we omit to record the changes to the configuration on the rest of the lattice.
	\begin{widetext}
	\begin{equation}\label{reftri}
	\raisebox{-0.5\height}{\includegraphics{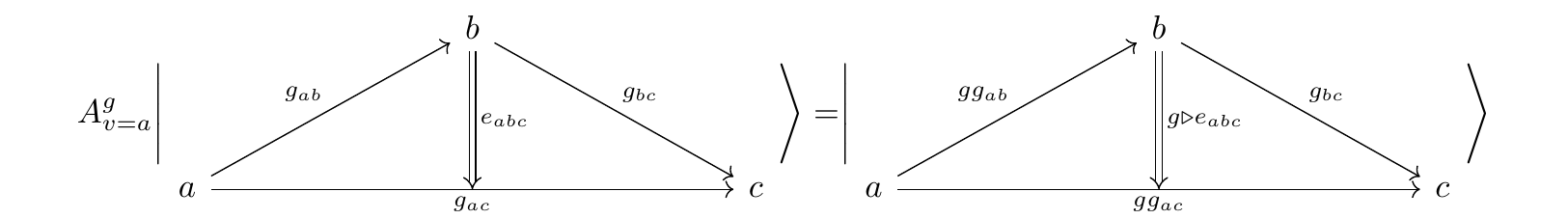}}
	\end{equation}
	
	\begin{equation}
	\raisebox{-0.5\height}{\includegraphics{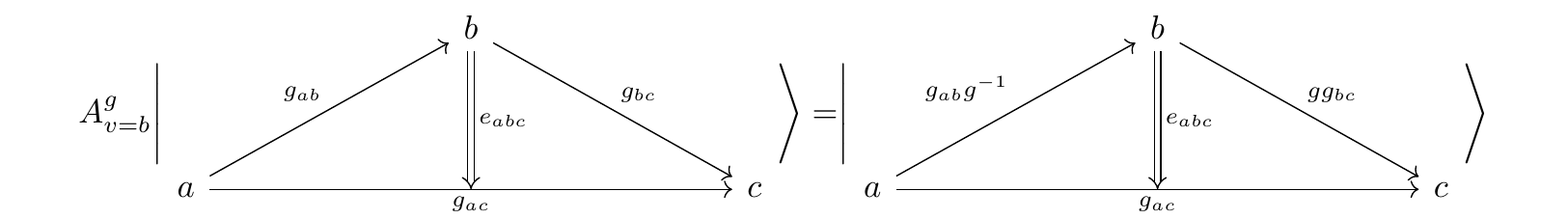}}
	\end{equation}
	\begin{equation}
	\raisebox{-0.5\height}{\includegraphics{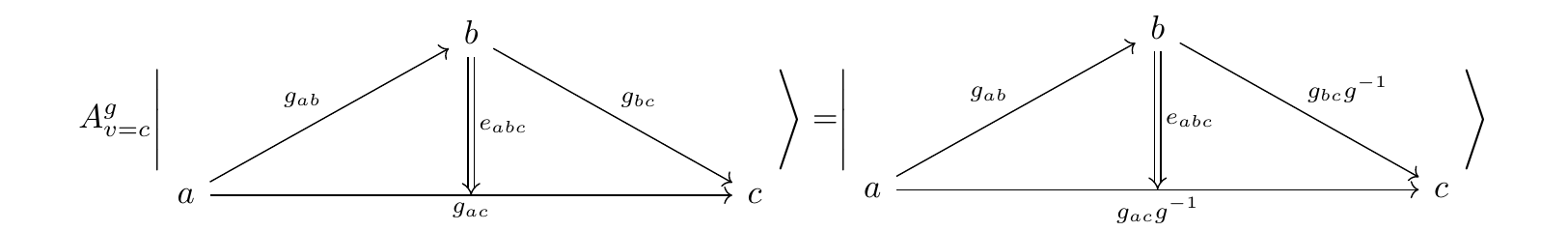}}
	\end{equation}
	\begin{equation}
	\raisebox{-0.5\height}{\includegraphics{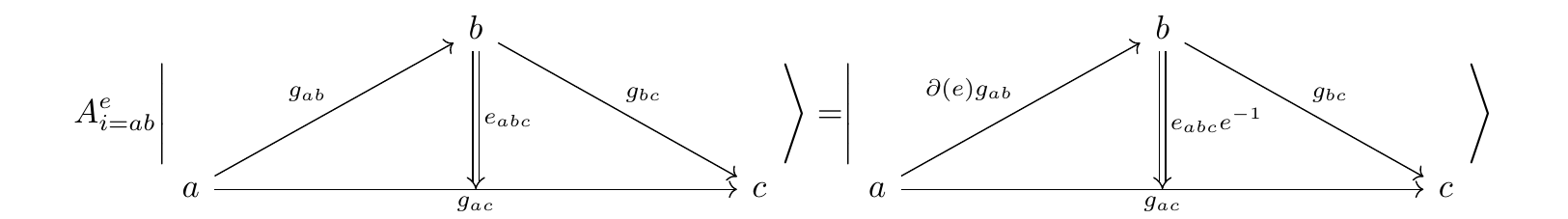}}
	\end{equation}
	\begin{equation}
	\raisebox{-0.5\height}{\includegraphics{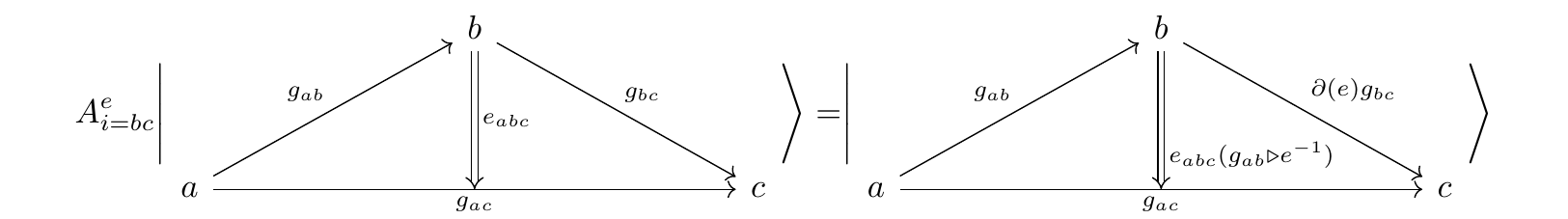}}
	\end{equation}
	\begin{equation}
	\raisebox{-0.5\height}{\includegraphics{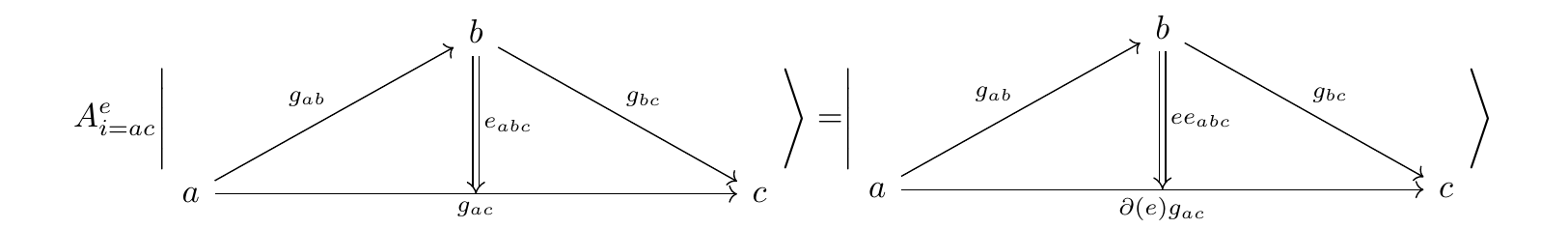}}
	\end{equation}
	\end{widetext}
	Now we can compute the transformations of the 1- and 2-holonomy under all possible gauge transformations of the triangle 
	(three vertex gauge transformations and three edge gauge transformations) and the tetrahedron 
	(four vertex gauge transformation and six edge gauge transformations). The formulas are given in Appendix \ref{b}.
	An immediate corollary of them is that denoting the
	transformed quantities by tilde we have
	\begin{equation}
	H_1(p)= \One \;\;\Longleftrightarrow\;\; \tilde{H}_1(p)=\One, p\in L^2\label{1fflat}
	\end{equation}
	Another consequence is
	\begin{equation}[B_p,A^g_v]=[B_p,A^e_i]=0.\end{equation}
	By \eqref{1fflat}, operators $A_v^g,A_i^e$ restrict to elements of End$({\cal H})$
	(preserve $H_1(p)=1$ with $p\in L^2$).
	When restricted we also have
	\begin{equation}
	H_2(P)= \One \;\;\Longleftrightarrow \tilde{H}_2(P)= \One,
	P=\partial x, x\in L^3\end{equation}
	Thus, 
	when restricted to ${\cal H}$, 
	they are gauge transformations (transform covariantly). 
	
	Hereafter we understand the operators as restricted to ${\cal H}$. 
	Consequently
	the commutation relations 
	\begin{equation}[B_P,A^g_v]=[B_P,A^e_i]=0\label{com4}\end{equation}
	hold for all possible values of the parameters.
	\subsection{The Hamiltonian}
	
	Now we define the vertex operators $A_v$ and edge operators $A_i$ as
	
	\[ A_v = \frac{1}{|G|} \sum_{g \in G} A_v^g \, , \quad \quad  A_i = \frac{1}{|E|} \sum_{e \in E} A_i^e \, .\]
	Using relation (\ref{gr1}) we can verify that $A_v$ is a projector operator, that is $A_v^2= A_v$:
        \begin{eqnarray*}A_v^2&=&\frac{1}{|G|^2} \sum_{g,h \in G} A_v^g A_v^h = \frac{1}{|G|^2} \sum_{g,h \in G} A_v^{gh} =\\ 
        &&\frac{1}{|G|^2} \sum_{g',h \in G} A_v^{g'} =\frac{1}{|G|} \sum_{g' \in G} A_v^{g'} = A_v.\end{eqnarray*}
	Similarly, using (\ref{gr2}) one obtains $A_i^2 = A_i$. Also, they satisfy the following commutation relations
	\begin{align}
	&[A_v,A_i]=0  \label{com1}\\
	&[A_v,A_{v'}]=0   \label{com2}\\
	&[A_i,A_{i'}]=0 \label{com3}
	\end{align}
	for any vertices $v,v'$ and edges $i,i'$. In fact, the relations (\ref{com2}) and (\ref{com3}) follow immediately from relations (\ref{nb1}) and (\ref{nb2}), respectively. In the same way, (\ref{com1}) follows from (\ref{mixe}) if $v \neq s(i)$. If $v= s(i)$, using (\ref{mixh}) we obtain
	
	\[ \sum_{e\in E, g\in G} A_v^g A_i^e  = \sum_{e\in E, g\in G} A_i^{g \trr e} A_v^g  =  \sum_{e'\in E, g\in G} A_i^{e'} A_v^g ,\]
	where in the last equality we used the fact that the map $g \trr (\cdot)$ is a bijection. Then it is clear that (\ref{com1}) holds.
	Let us now consider the multiplicative operators $B_P=\delta_{H^2(P),{\bf 1}}$ defined above. It is clear that they mutually 
	commute and they are projections. It is also clear that they commute with the projections $A_v$ and $A_i$ due to (\ref{com4}).
	Now we can write down a {\em Hamiltonian of higher lattice gauge theory} in terms of mutually commuting operators in End$({\cal H})$
	\begin{equation}H=-\sum_v A_v-\sum_i A_i-\sum_P B_P\label{HBull}\end{equation}
	where the summations run over vertices $v\in L^0$, edges $i\in L^1$ and blob boundaries $P=\partial x, x\in L^3$.

	Note that one can define
	a Hamiltonian on ${\cal H}^L$ with the same groundstate sector by including the 1-flatness constraint operators:
	\begin{equation}H'=-\sum_v A_v-\sum_i A_i-\sum_P B_P(\prod_p B_p)-\sum_p B_p
	\label{Marcos}
	\end{equation}
	where summation and product over $p$ mean over elements $p\in L^2$. The reason for the modified form of 
	the $B_P$ operators is that without the multiplier that enforces fake-flatness, they would not commute with the gauge transformations
	(as can be seen from the transformation of $H_2(P)$ under $A_{cd}^e$ of the tetrahedron $P=a<b<c<d$), see Appendix \ref{b}.
	This way $H'$ is also a sum of mutually commuting projections and the groundstates of $H$ and $H'$ agree.
	
	In the restriction to a two-manifold and crossed module $\mathcal{G}=(G,\One_{G},\partial,\trr)$ we note the model $H'$ reproduces the Kitaev quantum double\cite{kitaev2003fault} Hamiltonian for group $G$. The second term is zero due to the triviality of $E$ and the third term does not enter the equation as there are no blobs in $d=2$, so we have
	\begin{equation}
	H=-\sum_{v}A_v-\sum_{p} B_p\ .
	\label{QDHami}
	\end{equation}
	From this connection, the first term in \eqref{HBull} can be seen naively as the Gauss constraint and the last term as the magnetic constraint. Caveat: Unless fake-flatness constraints are imposed, the 2-holonomy of a blob is ambiguous: the formula \eqref{2hol} depends on the choice of the composition of the 2-holonomies of the boundary faces of the blob\cite{pfeiffer2003higher}.
\section{Relation to the 4D Yetter TQFT\label{Yetterrelation}}
\subsection{Ground state projection as a 4D state sum}
In this chapter we will relate our Hamiltonian model in three space dimensions to the Yetter TQFT in four dimensions. The fact that one
can associate a $d$-dimensional Hamiltonian lattice model to a $D=d+1$
dimensional TQFTs is not new.
For $d=2$ K\'ad\'ar et 
al. showed \cite{KMR08} that 
the Levin-Wen model \cite{Stringnet} is the Hamiltonian version of the Turaev-Viro (TV) 
TQFT \cite{TV92,BW1996} in that the groundstate projection 
of the former defined on the two dimensional lattice $L$ for $M^2$ is given by the TV state sum for the $M^2\times [0,1]$. 
The rigorous proof was 
given by Kirillov \cite{kirillov2011string}. 

A subset of the Levin-Wen Hamiltonian models was shown\cite{KMR09,BA09} to be the 
the dual lattice description of those corresponding to BF gauge theories: the Quantum Double models of Kitaev\cite{kitaev2003fault}.
Here duality means Fourier transformation on the gauge group, as a result of which states are labeled by irreducible 
representations instead of group elements. Hence, the above statement has to hold in the dual description as well. We will state and 
prove it in this section and generalise to the case of our 3D model: its ground state projection is given by the appropriate 4D 
Yetter TQFT amplitude.

The general correspondence we will investigate has been known qualitatively.   
The 2+1 BF-theory action with gauge group $SU(2)$ is equivalent to the Einstein-Hilbert one for Euclidean signature 
and zero cosmological constant as shown by e.g., Ooguri and Sasakura\cite{OS}. On the other hand, in 1969, Ponzano and Regge derived the gravity
action from the asymptotic form of the Wigner-Racah coefficients \cite{PR}. These results were the motivation for several works in 
the 2+1 quantum gravity literature, where the details of the correspondence were well understood for the case of $SU(2)$ \cite{FL1}: the
Hilbert space is the state space of canonical quantum gravity, the TQFT is the corresponding state sum or spin foam model.   

We will sketch the derivation of the TQFT state sum from a general BF gauge theory. Then we replace the continuous gauge group with
a finite one and work out the correspondence for $d=2,3$ for lattice and higher lattice gauge theory (LGT, HLGT) and show the pattern, which
arises for arbitrary dimension.  

\subsection{Ordinary Pure Lattice gauge theory based on a finite group}
Let us first consider the theory defined on an oriented manifold $M^D$ of dimension $D$ 
by the action with a compact Lie group $G$ and its Lie algebra ${\mathfrak g}$
\[ S[B,A]=\int_{M^D}\mbox{tr}\left(B\wedge F(A)\right)\]
with $F(A)=dA+[A,A]$ being the (${\mathfrak g}$-valued) curvature of the connection $A$;
$B$ a locally ${\mathfrak g}$ valued (D-2)-form; and
tr 
the Cartan-Killing form of ${\mathfrak g}$.
The partition function (a map that associates a scalar to each manifold $M^D$) 
is defined formally in terms of the path integral
\[Z_{BF}(M^D)=\int {\cal D}B{\cal D}A\,e^{iS[B,A]}\]
with ${\cal D}A$ and ${\cal D}B$ standing for some measure over the space of connections and the $B$-field. 
To make sense of this formal expression there is a standard discretization 
procedure, see e.g., Oeckl \cite{Oeckl} or Baez\cite{BaezBF}.
Here we will only sketch the procedure and write down the 
discrete version of the partition function.
Let $\Delta$ be 
a dressed lattice of $M^D$ and $\tilde{\Delta}$ the dual complex
(whose $k\leq D$ dimensional simplices are in 
one-to-one correspondence with the $D-k$ simplices of $\Delta$).
We  dress $\tilde{\Delta}$ similarly to before:
we orient edges
(which are dual to $(D-1)$-simplices of $\Delta$)
and give circular orientation to faces and 
distinguish basepoints in each face.
A gauge configuration is an assignment of a group element to each edge.
For a face $p$ we define the exponentiated curvature by $g_p=\prod_{i\in \partial p}  g_i$: 
the multiplication is done in the order along the chosen cyclic orientation starting at $v_0(p)$.   
The $B$-field, locally being a $D-2$ form,
is naturally associated to dual faces.
$F_p$ is the curvature variable associated
to the dual face $p$.
Now, the integral $\int dB_p e^{iB_p F_p}$ vanishes, unless the curvature vanishes, thus it can be replaced by 
$\delta_{g_p,{\bf 1}}=\delta_{g_p,{\bf 1}}$ 
in terms of the chosen variables: the vanishing of the curvature in the dual 
face $p$ is equivalent to the trivial holonomy along the boundary of the face. 
The expression for the discretized path 
integral reads
\begin{equation}
Z_{LGT}(M^D,\Delta)=\int\prod_{i\in \Delta^1}dg_i\prod_{p\in \Delta^2} \delta_{g_p,{\bf 1}}\ ,\label{partf}
\end{equation}
where $\Delta^1$ ($\Delta^2$) is the set of edges (faces) of
$\tilde{\Delta}$
(see \cite{Oeckl,FLd,FL1,BaezBF} for details of the
discretization procedure).
For compact Lie groups the measure $dg_i$ is the Haar-measure on $G$.  
The partition function is the sum of all colouring subject to the constraint that the holonomy
$g_p$ around each face is trivial: the underlying connection is flat. 

At this point there is no difference between 
using $\Delta$ or $\tilde{\Delta}$ for edge and face coloring. We will use the former.

Note that we still do not know whether 
$Z_{LGT}<\infty$ in general,
but we are interested here in replacing $G$ with a finite group\footnote{The most studied case of a compact Lie
	group is $SU(2)$, and finiteness requires gauge fixing \cite{FL1}}. 
For a finite group $G$, we will choose the measure $\int dg=\sum_{g\in G}$. 
The partition function needs to be normalized so that it is independent of the lattice used:
\begin{equation}
Z_{LGT}(M^D,\Delta)=|G|^{-|\Delta^0|}|\{\mbox{{\em admissible} colourings of}\,\Delta\}\label{zlgt}|\ ,
\end{equation} 
where admissibility means that all faces are flat: $\prod_{p\in \Delta^2}\delta_{g_p,{\bf 1}}=1$ 
and the multiplicative factor ensures independence on
the lattice. This formula holds in arbitrary dimension (it is proved in more general settings by 
Porter\cite{porter1998topological}, Yetter\cite{yetter1993tqft} and Faria Martins / Porter\cite{martins2007yetter}).  

Generalising to manifolds with boundary, we can seek to satisfy the
(weak)
cobordism property of a TQFT
(partition functions become partition vectors with composition by
dot product --- see e.g. Martin\cite[\S2.1,\S10.2]{Martin91}).
For this 
the above definition has to be  
generalised. 
Let the boundary be a closed $d=D-1$ dimensional 
manifold with a lattice $\Delta^{(b)}\subset\Delta$.
A boundary colouring is denoted by 
$\Delta^{(b)}_c$ (so $c$ stands for a set $\{g_i\in
G,i\in \Delta^{(b)1}\}$).
Let us call a colouring of $\Delta$ $c$-admissible 
if it restricts to $\Delta^{(b)}_c$
and if the flatness constraints for all faces are satisfied: 
$\prod_{p\in \Delta^2}\delta_{g_p,{\bf 1}}=1$.
The partition vector is a vector with components indexed by
the possible boundary colourings $\Delta^{(b)}_c$, 
and a component then reads
\begin{equation}\begin{array}{l}
Z_{LGT}(M^D,\Delta,\Delta^{(b)}_c)=|G|^{-|\Delta^0|+\frac{|\Delta^{(b)0}|}{2}}|\\\\\{\mbox{c-admissible colourings of}\,\Delta\}|
\end{array}\label{bzlgt}\end{equation}
This way the cobordism property of a TQFTs holds as follows.
Let $N^d$ be a closed submanifold of a manifold $M^D$,
such that $N^d$ separates
$M^D$ into two unconnected components. Then write  $M^D_1$ and $M^D_2$
for the closures of the two components of $M^D \setminus N^d$. 
The manifolds $M^D_1$ and $M^D_2$ (with lattices $\Delta^1, \Delta^2$, say)
have homeomorphic and oppositely oriented boundaries $N^d$
(with lattice $L\subset \Delta_1,\Delta_2$). 
Then the partition functions satisfy the cobordism property of
TQFTs.
That is, changing the
notation as $\int\prod_{i\in L^1}dg_i\to \sum_c$ we can write
\[ \begin{array}{l}Z_{LGT}\left((M_1^D\sqcup_{N^d} M_2^D),\Delta_1\cup\Delta_2\right)=\\\\
\sum'_c Z_{LGT}(M_1^D,\Delta_1,L_c)Z_{LGT}(M_2^D,\Delta_2,L_c)\end{array}\]
where $M_1\sqcup_{N^d} M_2$ denotes $M^D$
(we think of it here as obtained from gluing $M_1$ and $M_2$ along $N^d$)
and the prime in 
${\displaystyle \sum'_c}$
indicates that flatness constraints have to be inserted for all $p\in L^2$. 

\subsubsection{2+1D lattice gauge theory with a finite group}
Recall the 2+1D Kitaev QD Hamiltonian from \eqref{QDHami}. The ground state projection reads 
\[{\cal P}^K_{gs}=\prod_{v\in L^0}A_v\prod_{p\in L^2}B_p.\]
\begin{theorem}
	Let $M^2$ be a 2-manifold with lattice $L$ and let $\Delta$ be the 3-dimensional lattice of $M^2\times [0,1]$ that restricts to
	$L_0\simeq L_1\simeq L$ at the boundaries $M^2\times \{0\}$ and $M^2\times \{1\}$. Let the internal edge set be 
	$\{v\times [0,1]\}_{v\in L^0}$. Let $L^i_j$ refer to $L_j$ and $L_{jc}$ the edge colorings of $L^1_j$. 
	Finally, we consider the QD Hilbert space based on $L_j$ with states $|L_{jc}\rangle$ and identify 
	the Hilbert spaces based on $L_j, j\in {0,1}$. Then 
	\begin{equation}Z_{LGT}(M^2\times [0,1],\Delta,L_{0c}\cup L_{1c})=
	\langle L_{1c}|{\cal P}^K_{gs}|L_{0c}\rangle\label{lgt}\end{equation}
	\label{th1}
\end{theorem}
The proof is written in Appendix \ref{c}.
Note, that we made a choice for using the edge and face set of $\Delta$ for defining the partition function. An 
alternative approach, more standard in the realm of the Turaev-Viro model \cite{BKET,KMR08,BaezBF}, makes use of the edge and face set of the dual lattice $\tilde{\Delta}$. 
Then, the minimal lattice, which restricts to $L^j$ at the boundary 
is three translated copies of $L$ connected with 
vertical edges\footnote{One considers $\tilde{L}\times [0,1]$ and constructs the dual complex of this. It will have a
	vertex in the middle of each prism connected vertically to the middle points of $\tilde{L}_j$, middle points of neighbor
	prisms are connected and the duals of $\tilde{L_j}$ are the original graphs $L_j$ at the boundary.}. 
That way, no Dirac-deltas are needed for boundary faces as those in the middle layer of $L$ enforce 
flatness for the gauge equivalent boundaries $L_{jc}$ and the proposition is stated identically to the Fourier dual
(Turaev-Viro vs. Levin-Wen) case.

\subsection{Higher lattice gauge theory based on a finite crossed module}
Fix now a crossed module $(G,E,\partial, \trr)$ and a four manifold $M^4$. 
Let us consider the theory given by the $BFCG$ action \cite{GPP}
\[ S[A,B,C,\Sigma]=\int_{M^4}\left(\mbox{tr}_{\mathfrak g}(B\wedge F_A)+\mbox{tr}_{\mathfrak h}(C\wedge G_\Sigma)\right)\]	
where $B$ is a $G$-valued 2-form, $F_A=dA+[A,A]$ is the curvature of the connection $A$, $C$ is an $E$-valued 1-form and
$G_\Sigma=d\Sigma+A\trr\Sigma$ the curvature 3-form of the two-connection $\Sigma$ corresponding to the gauge group $E$.
We will use the following form of the partition function  
\[Z_{BFCG}(M^{4})=\int {\cal D}A\,{\cal D}B\,{\cal D}C\, {\cal D}G\,e^{iS[A,B,C,G]}\]
whose discretized form defined on the dressed lattice $\Delta$ of $M^4$ is given by
\[Z_{HGT}(M^{4})=\int\prod_{i\in L^1}dg_i\prod_{p\in L^2}de_p\,\delta_{H_1(p),{\bf 1}}\prod_{t\in L^3}\delta_{H_2(t),{\bf 1}}.\]
For a finite group we can rewrite it analogously to (\ref{zlgt}) in LGT. Let a colouring be called admissible 
if all Dirac-delta constraints are satisfied: all faces are fake-flat and all blobs are 2-flat. We do the substitutions
$\int dg_i\to \sum_{g_i\in G}$ and $\int de_p\to \sum_{e_p\in E}$ and write
\[Z_{HGT}(M^{4})=\frac{|E|^{|\Delta^0|-|\Delta^1|}}{|G|^{|\Delta^0|}}|\{\mbox{admissible colourings of}\, \Delta\}|\ ,\]
where the multiplicative factor ensures independence on the lattice. This is proved to be the same in arbitrary 
dimensions\cite{porter1998topological,yetter1993tqft,martins2007yetter}.
For manifolds with boundary, we need to modify the above similarly to the LGT case. Let the manifold $M^4$ with boundary 
have a lattice decomposition $\Delta$, and let $\Delta^{(b)}\subset \Delta$ denote the boundary lattice. Let $\Delta^{(b)}_c$ be a colouring of $\Delta^{(b)}$:
$\{g_i\in G,i\in \Delta^{(b)1}, e_p\in E, p\in \Delta^{(b)2}\}$. We call a colouring of $\Delta$ c-admissible if it is 
admissible in $\Delta$ and restricts to $\Delta^{(b)}_c$ on
$\Delta^{(b)}$. The components of the partition vector read:
\begin{equation}\begin{array}{l}
Z_{HGT}(M^4,\Delta,\Delta^{(b)}_c)=
\frac{|E|^{|\Delta^0|-|\Delta^1|}}{|G|^{|\Delta^0|}}\,
\frac{|G|^{\frac{|\Delta^{(b)0}|}{2}}}{|E|^{\frac{|\Delta^{(b)0}|-|\Delta^{(b)1}|}{2}}}\times
\\\\|\{\mbox{c-admissible colourings of}\,\Delta\}|
\end{array}\label{bzhlgt}\end{equation}
This does not depend on the lattice decomposition of $M^4$ extending a given lattice decomposition of the boundary \cite{BCKMM2}.

\begin{theorem}
	\label{th2}
	Let $\Delta_L\equiv \Delta$ be the lattice of $M^3\times [0,1]$ which restricts to
	$L_0\simeq L_1\simeq L$ at boundaries $M^3\times \{0\}$ and $M^3\times \{1\}$ and the internal edge set is 
	$\{v\times [0,1]\}_{v\in L^0}$. Let $L^i_j$ refer to $L_j$ and $L_{jc}$ the colorings of those, where fake flatness
	is assumed for each face $p\in L^2_j,j=1,2$. 
	Finally, we consider the Hilbert space ${\cal H}$ defined in section \ref{Hami} based on $L$ 
	with states $|L_c\rangle$ and identify 
	the Hilbert spaces based on $L_j, j\in {0,1}$ with it and consider the projection 
	${\cal P}^B_{gs}$ to the groundstate defined by
	\[{\cal P}^B_{gs}=\prod_{v\in L^0}A_v\prod_{i\in L^1}A_i\prod_{P\in L^3} B_P\] 
	Then
	\begin{equation}
	Z_{HGT}(M^3\times [0,1],L_{0c},L_{1c})
	=  \langle L_{1c}|{\cal P}^B_{gs}|L_{0c}\rangle\ .
	\label{hlgt}\end{equation}
\end{theorem}
In words, the groundstate projection of our 3D Hamiltonian model associated to $M^3$ is given by the Yetter TQFT amplitude on 
$M^3\times [0,1]$.
The proof is in Appendix \ref{d}.
\subsection{Hamiltonians corresponding to lattice gauge theories}
We will look at the correspondence in arbitrary dimension $d\geq 1$ for both ordinary and higher lattice gauge theories. 
We can observe the following.
\begin{itemize}
	\item The fake flatness constraints of internal faces and 2-flatness of internal blobs 
	of the $d+1$ dimensional prism lattice are equivalent to the 
	bottom and top layer of the prism being connected by gauge transformations. For ordinary gauge theory, the latter
	is equivalent to the flatness of internal faces. 
	\item The 2-flatness (flatness) constraints of internal faces of the boundary lattice are the magnetic operators 
	of the Hamiltonian (in ordinary gauge theory, respectively). 
\end{itemize}
As a consequence, the Hamiltonian in the Hilbert space associated to the $d$ dimensional lattice $L$, 
whose groundstate projections are given by the corresponding $d+1$ dimensional partition function is given by
the following table, where the sign $-||-$ means the same formula as on its left for all terms to the right starting from
the sign\footnote{The Hamiltonian for $d=3$ HLGT in the table differs from (\ref{HBull}) by an unimportant constant.}.
\begin{widetext}
\[\begin{array}{c|c|c|c|c|c}d&1&2&3&&\\\hline
LGT&{\displaystyle-\sum_{v\in L^0} A_v}&{\displaystyle-\sum_{v\in L^0} A_v-\sum_{p\in L^2}}B_p&-||-&\dots\\\hline
HLGT&{\displaystyle-\sum_{v\in L^0} A_v-\sum_{i\in L^1} A_i}&{\displaystyle-\sum_{v\in L^0} A_v-\sum_{i\in L^1} A_i}
&{\displaystyle-\sum_{v\in L^0} A_v-\sum_{i\in L^1} A_i-\sum_{P\in L^3}}B_P&-||-&\dots
\end{array}\]
\end{widetext}

\subsection{The ground state degeneracy\label{gsdsec}}
We compute here the groundstate degeneracy (GSD) for a few examples. This is given by the trace of the 
groundstate projection. By virtue of the theorems, is can also be computed from the invariant
\begin{equation}\label{hgsd}\begin{array}{r}
\mbox{Tr}(P_{gs}^B)=Z_{HGT}(M^d\times S^1)=|G|^{-|L^0|}|E|^{-|L^1|}\times\\\\|\{\mbox{admissible colouring of}\,\Delta_L\}|
\end{array}\end{equation}
with $L$ being the lattice of $M^d$ and $L^i$ referring to its set of $i$-dimensional cells as before and $\Delta_L$ is
the prism lattice whose top $L_1$ and bottom $L_0$ are identified.
It applies to LGT too with the obvious modifications. 
\begin{itemize}
	\item $d=1$.  Here a minimal lattice of $S_1$ is the lattice with one edge with
	its source and target vertex identified. Let us first consider the case of ordinary lattice gauge theory; i.e $E$ is the trivial group. The GSD is by definition is the number of gauge equivalence classes of admissible colourings of $S^1$. This clearly coincides with the number of   
	of conjugacy classes of $G$. We can also obtain this GSD as 
	$Z_{LGT}(S^1\times S^1)$, which is $|G|^{-1}$ times the number of colourings
	of the lattice of the torus; explicitly: $|G|^{-1} \,|\{\phi: \pi_1(T^2)\to G\}|=|\{(g,h) \in G^2\colon gh=hg\}$. So in this case, the equality \eqref{hgsd} boils down
	to the well established fact from group theory that the number of conjugacy classes of a finite group equals the order of the group times its commuting fraction, the probability that two elements commute; i.e. $\frac{1}{|G|}\, | \{(g,h)\in G^2\colon gh=hg \}|=\textrm{number of conjugacy classes of } G.$
	
	In the general HGT case and also for $S^1$, looking at the lhs of \eqref{hgsd}, the GSD can be expressed as the number of conjugacy classes of $G/\partial(E)$. The rhs. of \eqref{hgsd} explicitly is: $\frac{1}{|G|\,|E|}|\{(g,h,e) \in G \times G \times E\colon \partial(e)=[g,h]\}|$. It is amusing the check these two coincide, which follows from the group theory fact stated in the above paragraph.  
	
	\item $d=2$.
	Here, the ordinary gauge theory model is well studied, the dimension 
	of the groundstate for the manifold $M^2$ and gauge group $G$ is $|\{\phi:\pi_1(M^2)\to G\}/\sim|$, where $\sim$ means modulo
	an overall conjugation $\phi(g)\mapsto h\phi(g)h^{-1}, g\in \pi_1(M^2), h\in G$. 
	That is, the GSD is the number of gauge equivalent classes of flat 
	connections. For $M^2=T^2$, this is well known to coincide with the number of 
	irreps of the double $DG$. Computing the GSD from the partition function gives 
	$\frac{1}{|G|}|\{(g_1,g_2,g_3),\in G^3, [g_i,g_j]={\bf 1}\}|$. Recalling that the irreps of the DG are in one-to-one 
	correspondence of the irreps of the centraliser subgroups of the representatives of conjugacy classes of $G$, the equality
	is clear. 
	
	For HGT, consider  $S^2$, with a cell decomposition with one single vertex and a unique 2-cells. (This would be a trivial case to consider for ordinary gauge theory.) In this case the GSD can be computed, looking at the lhs of \eqref{hgsd} as being the cardinality of the set of orbits of the action of $G$ on $\ker(\partial)$. Computing the GSD from the rhs of \eqref{hgsd} yields 
	$\frac{1}{|G|}\sum_{e \in \ker(\partial)} |\{g \in G\colon g \trr e=e\}|$. Elementary tools from group actions tell us that these two coincide.   
	
	\item $d=3$. Let us consider the $T^3$ case, using the obvious cell decomposition of the cube, and then identifying sides. 
	Fake flatness of the three distinct faces and 2-flatness of the cube read ($[g_1,g_2]\equiv g_1g_2g_1^{-1}g_2^{-1}, g_1,g_2\in G$): 
	\[\begin{array}{c}[x,y]=\partial f,\quad [x,z]=\partial e,\quad[y,z]=\partial k,\\\\f\, (yxy^{-1} \trr k^{-1}) \, (y \trr e) \, (yzy^{-1} \trr f^{-1})\, k=e\end{array}\]
	Let the subset of $G^3 \times E^3$ defined by the joint solutions 
	of the above equations be denoted by $S$ and consider the equivalence relation $\cong$ in $S$  generated by
        (with $a \in G$ and $e_x,e_y,e_z \in E$):
	\[\begin{array}{l}\quad\,\big(x,y,z,e,f,k\big)\\\\
	\cong\big(axa^{-1},aya^{-1},aza^{-1},a \trr e, a \trr  f, a \trr k\big)\ ,\\\\
	\cong \left (x,\partial e_y\, y,z,e, (x \trr e_y)\, f\, e_y^{-1},e_y \,k\, (z \trr e_y^{-1})    \right)\ ,\\\\
	\cong \left (x,y,\partial e_z\, z,(y \trr e_z)\,e\,e_z^{-1},f,(x \trr e_z)\, k\, e_z^{-1}  \right)\ ,\\\\
	\cong \left (\partial e_x\,x,y,z,e_x\, e \, (z \trr e_x^{-1}), e_x \, f\, (y \trr e_x^{-1})\, k  \right).
	\end{array}\]
	The GSD is $|S/\cong|$. For instance consider the crossed module $DG=(G,G,\mbox{id},\mbox{ad})$, 
	(where ad stands for the conjugation action). Here GSD$\,=1$, and this is easily computable from the rhs. of (\ref{hgsd}).

	Another easily computable example is $({\mathbb Z}_2,{\mathbb Z}_{2r},\mbox{sgn},\mbox{id})$, where ${\rm sgn}$ is the parity 
	($r\in {\mathbb N_+}$), and $id$ denotes the trivial action. Here 
	GSD$\,=r^3$. 
	This is easily seen from the rhs. of (\ref{hgsd}): $x,y,z\in G$ are
	arbitrary, they determine $(e,f,k)$ via the fake flatness constraints and the 2-flatness of the cube holds by construction. 
	We have three more cubes based on 2- and 3-faces, whose faces are again pairwise identified. The identical argument 
	applies: the new face labels are determined by the fake flatness of the sides, so the four edges labels are arbitrary and
	all 6 face labels are determined by the commutators. So we have $|G|^4$ admissible colourings, $|L^0|=1$ and $|L_1|=3$.
	Another way to infer that ${\rm GSD}=r^3$ is the following. Yetter's state sum $Z_{HGT}(W)$, where $W$ is a closed manifold, depends only on the weak homotopy type of the underlying crossed module\cite{martins2007yetter}. The crossed module $({\mathbb Z}_2,{\mathbb Z}_{2r},\mbox{sgn},id)$ is weak equivalent to $(\{0\},{\mathbb Z}_{r})$, where we consider the constant map $\partial\colon {\mathbb Z}_{r}\to \{0\}$. In general, considering a crossed module of the form $\mathcal{E}=(\mathbf{1}_{E},E,\partial,\trr)$, where $\partial$ and $\trr$ are trivial maps, we have that $Z_{HGT}(S^1\times S^1 \times S^1 \times S^1)=|E|^3$. The number of admissible colouring is $|E|^4$ since we can colour the 2-cells of the 4-cube with faces identified as we please. 
\end{itemize}

\section{Relation to Walker Wang Models\label{WalkerWang}}

In this section we discuss the relation between the Walker-Wang model\cite{WalkerWang} and our model. In particular we outline a duality map between our model with the finite crossed module $\mathcal{E}=(\mathbf{1}_{E},E,\partial,\trr)$, where $\partial:E\rightarrow \mathbb{1}_{E}$ and $\trr$ is the identity and the Walker-Wang model based on the symmetric fusion category $\mathcal{M}(\mathcal{E})$, where $E$ is any finite Abelian group.

\subsection{Walker-Wang Model}

To begin, we briefly outline the Walker-Wang model\cite{WalkerWang}. The Walker-Wang model is a 3+1D model of string-net condensation with groundstates proposed to describe time-reversal invariant topological phases of matter in the bulk and chiral anyon theories on the boundary\cite{Simon}. Such models are believed to be the Hamiltonian realisation of the Crane-Yetter-Kauffman TQFT\cite{crane1993categorical} state sum models analogous to the relation between our model and the Yetter's homotopy 2-type TQFT\cite{yetter1993tqft}.

\begin{figure}
	\centering
	\begin{tikzpicture}[baseline={([yshift=-25pt]current bounding box.north)}]
	\def\z{2}
	\draw[] (0,0) to (0,0.5*\z);
	\draw[] (0,0) to (0,-0.5*\z);
	\draw[] (0,0) to (-0.5*\z,0);
	\draw[] (0,0) to (0.5*\z,0);
	\draw[] (0,0) to (0.25*\z,0.25*\z);
	\draw[] (0,0) to (-0.25*\z,-0.25*\z);
	\end{tikzpicture}\hspace{.5cm}
	\begin{tikzpicture}[baseline={([yshift=-0pt]current bounding box.north)}]
	\draw[->] (0,0) to (1,0);
	\end{tikzpicture}\hspace{.5cm}
	\begin{tikzpicture}[baseline={([yshift=-25pt]current bounding box.north)}]
	\def\z{2};
	\def\o{0.2};
	\draw[] (-0.5*\z,0) to (0.5*\z,0);
	\draw[] (0+\o,0) to (0+\o,-0.5*\z);
	\draw[] (0-\o,0) to (0+\o,\o);
	\draw[] (\o,\o) to (\o,0.5*\z);
	\draw[] (-2*\o,0) to (-0.25*\z-2*\o,-0.25*\z-\o);
	\draw[] (0+\o,\o) to (0.25*\z+2*\o,0.25*\z+\o);
	\end{tikzpicture}
	\caption{Resolution of 6-valent vertex to a trivalent vertex.}
	\label{vertexresolution}
\end{figure}
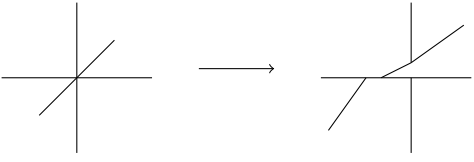

The Walker-Wang model is specified by two pieces of input data, a unitary braided fusion category (UBFC) $\mathcal{C}$ and a cubulation $C$ of a 3-manifold $M^3$. In the following we will define the generic model on a trivalent graph $\Gamma$ defined from the 1-skeleton $C^1$ of $C$ where vertices are canonically resolved to trivalent vertices see fig \ref{vertexresolution}. We will then restrict the input to a symmetric braided fusion category $\mathcal{\mathcal{E}}$ and remove the vertex resolution condition. We will make the assumption that the cubulation of the manifold is simple: namely all faces have 4-edges and each vertex is 6-valent\footnote{Every 3-manifold has a presentation in terms of a cubulation, in other words in terms of a partition into 3-dimensional cubes, which only intersect along a common face. However in some cases the valence of some of the edges of a cubulation may be different of 4, and therefore some vertices may not be six-valent. For some manifolds these features are not avoidable; see \cite{cooper1988triangulating}.}. Later on we will make the
additional assumption of working with cubulations such that the
dual cell decomposition also is a cubulation.

\begin{figure}[]
	\centering
	\includegraphics[width=\columnwidth]{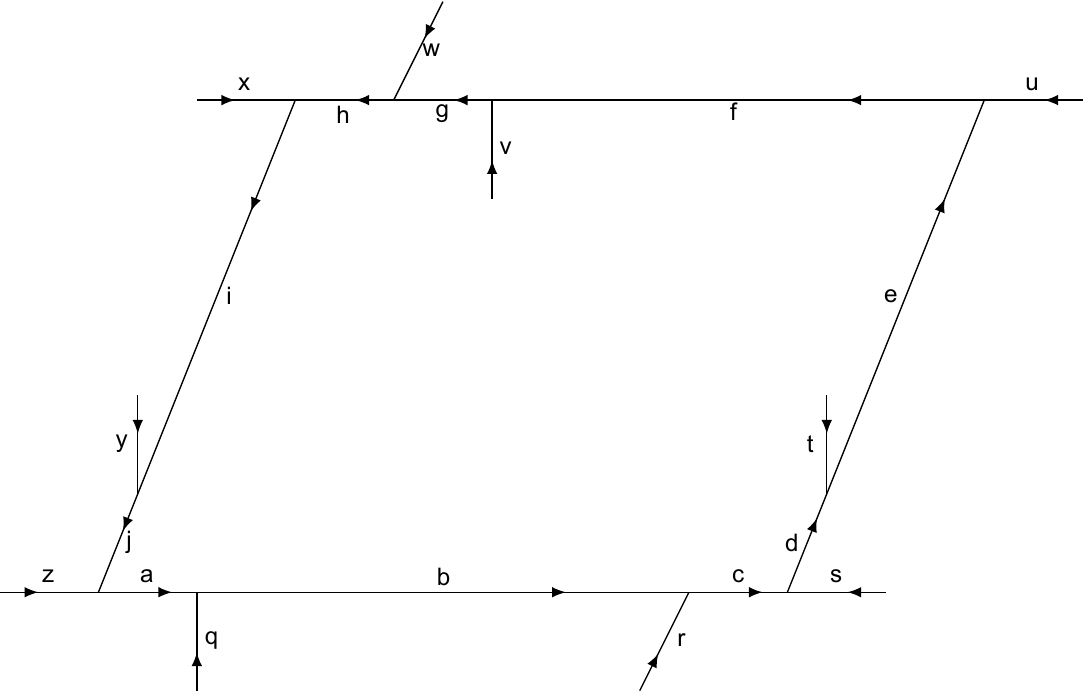}
	\caption{Trivalent plaquette with oriented edges for Walker-Wang model.}
	\label{WWp}
\end{figure}

The Walker-Wang model is defined on the trivalent cubic graph $\Gamma$ (see fig \ref{WWp}) with directed edges. The Hilbert space has an orthonormal basis given by colourings of the directed edges of $\Gamma$ by labels from $\mathcal{L}=\{1,a,b,c,\cdots\}$. For each edge label $a\in\mathcal{L}$ there is a conjugate label $a^*\in\mathcal{L}$ which satisfy the relation $a^{**}=a$. We define the states such that reversing the direction of an edge and conjugating the edge label gives the same state of the Hilbert space as the original configuration. The label set $\mathcal{L}$ has a unique element $1\in\mathcal{L}$ we call the vacuum which satisfies the relation $1=1^{*}$.

To specify the Hamiltonian we introduce the fusion algebra of the label set\cite{Bonderson,KitaevHoney,Wangtqc,PachosB}. A fusion rule is an associative, commutative product of labels such that for $a,b,c\in\mathcal{L}$, $a\otimes b=\sum_{c}N_{ab}^{c}c$. Here $N_{ab}^{c}\in\mathbb{Z}^+$ is a non-negative integer called the fusion multiplicity. In the following we will restrict to the case of "multiplicity free" which is the restriction $N_{ab}^{c}\in\{0,1\}$ $\forall a,b,c\in\mathcal{L}$. The fusion multiplicities satisfy the following relations
\begin{align}
N_{ab}^{c}&=N_{ba}^{c}\\
N^{1}_{ab}&=\delta_{ab^*}\\
N^{b}_{a1}&=\delta_{ab}\\
\sum_{x\in\mathcal{L}}N^{x}_{ab}N^{d}_{xc}&=\sum_{x\in\mathcal{L}}N^{d}_{ax}N^{x}_{cd}.
\end{align}

Given the label set and fusion algebra we define $d:\mathcal{L}\rightarrow\mathbf{R}$ such that $\forall a\in\mathcal{L}$, $d:a\mapsto d_a$ and $d_{a^*}=d_{a}$. We will refer to $d_a$ as the quantum dimension of the label $a$. The quantum dimensions are required to satisfy
\begin{equation}
d_ad_b=\sum_{c}N_{ab}^{c}d_c.
\end{equation}
Additionally we define $\alpha_{i}=sgn(d_{i})\in\{\pm 1\}$ which satisfies
\begin{equation}
\alpha_{i}\alpha_{j}\alpha_{k}=1
\end{equation}
if $N^{k^*}_{ij}=1$.

Given the fusion algebra and quantum dimensions we define the $6j$-symbols which enforce the associativity of fusion of processes. The $6j$-symbols are a map $F:\mathcal{L}^6\rightarrow \mathbf{C}$ which satisfy the following relations
\begin{align}
&F^{ijm}_{j^{*}i^{*}1}=\frac{v_{m}}{v_{i}v_{j}}N^{m^*}_{ij}\\
&F^{ijm}_{kln}=F^{klm^*}_{jin^*}=F^{jim}_{lkn^*}=F^{mij}_{nk^*l^*}\frac{v_{m}v_{n}}{v_{j}v_{l}}=\conjugatet{F^{j^*i^*m^*}_{l^*k^*n}}\\
&\sum_{n}F^{mlq}_{kp^*n}F^{jip}_{mns^*}F^{js^*n}_{lkr^*}=F^{jip}_{q^*kr}F^{riq^*}_{mls^*}\\
&\sum_{n}F^{mlq}_{kp^*n}F^{l^*m^*i^*}_{pk^*n}=\delta_{iq}\delta_{mlq}\delta_{k^*ip}
\end{align}  
where $v_{a}=\sqrt{d_a}$.

The final piece of data required to define the Walker-Wang model is the braiding relations or $R$-matrices. The $R$-matrices are a map $R:\mathcal{L}^3\rightarrow\mathbf{C}$ which are required to satisfy the Hexagon equations which ensure the compatibility of braiding and fusion. The Hexagon equations are as follows
\begin{align}
&\sum_{g}F^{cad^*}_{be^*g}R^{e}_{gc}F^{abg^*}_{ce^*f}=R^{d}_{ac}F^{acd^*}_{be^*f}R^{f}_{bc}\no
&\sum_{g}F^{e^*bd}_{cag}R^{e}_{ad}F^{e^*ag}_{bcf}=R^{d}_{ac}F^{e^*bd}_{acf}R^{f}_{ab}.
\end{align}
The data $(\mathcal{L},N,d,F,R)$ forms a UBFC. Examples of solutions to the above data are representations of a finite group or a quantum group.

Using the above data we can write down the Walker-Wang Hamiltonian. The Hamiltonian is of the following form
\begin{equation}
H=-\sum_{v\in\Gamma}A_v-\sum_{p\in\Gamma}B_p
\end{equation}
where $\Gamma$ is the directed, trivalent graph on which the model is defined and the $v$ and $p$ are the vertices and plaquettes of the graph. The plaquettes are defined with reference to the original square faces of $C$ before the vertex resolution. The term $A_{v}$ is the vertex operator and acts on the 3-edges adjacent to a vertex. We define the action of $A_v$ on states as follows
\begin{equation}
A_v\Biggr|
\begin{tikzpicture}[baseline={([yshift=-22pt]current bounding box.north)}]
\draw[middlearrow={latex}] (0,0) to (0,0.5);
\draw[middlearrow={latex reversed}] (0,0.5) to (0.25,1);
\draw[middlearrow={latex reversed}] (0,0.5) to (-0.25,1);
\node[font=\fontsize{7}{0}\selectfont] at (0.15,0.1) {$c$};
\node[font=\fontsize{7}{0}\selectfont] at (0.35,1.02) {$b$};
\node[font=\fontsize{7}{0}\selectfont] at (-0.35,1) {$a$};
\end{tikzpicture}
\Biggr\rangle=\delta_{abc}\Biggr|
\begin{tikzpicture}[baseline={([yshift=-22pt]current bounding box.north)}]
\draw[middlearrow={latex}] (0,0) to (0,0.5);
\draw[middlearrow={latex reversed}] (0,0.5) to (0.25,1);
\draw[middlearrow={latex reversed}] (0,0.5) to (-0.25,1);
\node[font=\fontsize{7}{0}\selectfont] at (0.15,0.1) {$c$};
\node[font=\fontsize{7}{0}\selectfont] at (0.35,1.02) {$b$};
\node[font=\fontsize{7}{0}\selectfont] at (-0.35,1) {$a$};
\end{tikzpicture}
\Biggr\rangle
\end{equation}
where $\delta_{abc}=1$ if $N^{c*}_{ab}\geq1$ and $\delta_{abc}=0$ else.

The plaquette operator $B_p$ has a slightly more complicated form in terms of the $6j$-symbols and $R$-matrices. Using Fig \ref{WWp} as the basis, $B_p$ has the following form
\begin{align}
B^n_p&=\sum_{a',b',c',d',e',f',g',h',i',j'}R_{t^*e}^{d}\conjugatet{R_{t^*e'}^{d'}}R_{v^*g'}^{f'}\conjugatet{R_{v^*g}^{f}}\nonumber\\
&F^{q b^* a}_{n^* a' b'^*}F^{r c^* b}_{n^* b' c'^*}F^{s d^* c}_{n^* c' d'^*}F^{t e^* d}_{n^* d' e'^*}F^{u f^* e}_{n^* e' f'^*}\nonumber\\
&F^{v g^* f}_{n^* f' g'^*}F^{w h^* g}_{n^* g' h'^*}F^{x i^* h}_{n^* h' i'^*}F^{y^* j^* i}_{n^* i' j'^*}F^{z^* a^* j}_{n^* j' a'^*}\nonumber\\
&\times\ket{a',b',c',d',e',f',g',h',i',j'}\bra{a,b,c,d,e,f,g,h,i,j}
\end{align}
\begin{equation}
B_p=\sum_{n\in\mathcal{L}} \frac{d_n}{D^2}B^n_p.
\end{equation}
We define the inner product of such states by
\begin{equation}
\braket{a,b,c,\cdots}{a',b',c',\cdots}=\delta_{aa'}\delta_{bb'}\delta_{cc'}\cdots.
\end{equation}

\subsection{The Symmetric Braided Fusion Category $\mathcal{M}(\mathcal{E})$}

Utilising the work of Bantay\cite{bantay2005characters} one can define a UBFC for every finite crossed module. Following this construction we will define the symmetric braided fusion category $\mathcal{M}(\mathcal{E})$ induced from the data of the finite crossed module $\mathcal{E}=(\mathbf{1}_E,E,\partial,\trr)$, where $E$ is any finite Abelian group and $\partial$ and $\trr$ are trivial.

The label set of $\mathcal{M}(\mathcal{E})$ is given by elements of $E$, with the vacuum label given by the identity element of $E$ and $a^*=a^{-1}$. The quantum dimension $d_{a}=1$ for all $a\in E$ and $D^2=|E|$. The fusion multiplicities are multiplicity free with $N_{ab}^{c}=\delta_{a+b,c}$ such that the fusion rules are given by the group composition rules (we use $+$ for the group composition as $E$ is an Abelian group) and $a\otimes b=a+b$ for all $a,b\in E$. We list the data of $\mathcal{M}(\mathcal{E})$ below. 
\begin{align}
\mathcal{L} &= \text{underlying set of $E$} \no
a\otimes b&=a+b \no
d_{a}&=1\qquad\forall a\in\mathcal{L}\no
D^2&=|E|\no
N^{c}_{ab}&=\delta_{a+b,c}\no
F^{ijm}_{kln}&=\delta_{i+j,m^{-1}}\delta_{k+l,m}\delta_{l+i,n^{-1}}\delta_{j+k,n}\no
R^{k}_{i+j}&=\delta_{i+j,k}	
\end{align}

\subsection{Walker-Wang Models for $\mathcal{M}(\mathcal{E})$}

Utilising $\mathcal{M}(\mathcal{E})$ as defined in the previous section as the input data of the Walker-Wang model we may write the terms of the Hamiltonian as follows. The vertex operator acts on basis elements as
\begin{align}
&A_v\Biggr|
\begin{tikzpicture}[baseline={([yshift=-22pt]current bounding box.north)}]
\draw[middlearrow={latex}] (0,0) to (0,0.5);
\draw[middlearrow={latex reversed}] (0,0.5) to (0.25,1);
\draw[middlearrow={latex reversed}] (0,0.5) to (-0.25,1);
\node[font=\fontsize{7}{0}\selectfont] at (0.15,0.1) {$c$};
\node[font=\fontsize{7}{0}\selectfont] at (0.35,1.02) {$b$};
\node[font=\fontsize{7}{0}\selectfont] at (-0.35,1) {$a$};
\end{tikzpicture}
\Biggr\rangle=\delta_{a+b+c,0}\Biggr|
\begin{tikzpicture}[baseline={([yshift=-22pt]current bounding box.north)}]
\draw[middlearrow={latex}] (0,0) to (0,0.5);
\draw[middlearrow={latex reversed}] (0,0.5) to (0.25,1);
\draw[middlearrow={latex reversed}] (0,0.5) to (-0.25,1);
\node[font=\fontsize{7}{0}\selectfont] at (0.15,0.1) {$c$};
\node[font=\fontsize{7}{0}\selectfont] at (0.35,1.02) {$b$};
\node[font=\fontsize{7}{0}\selectfont] at (-0.35,1) {$a$};
\end{tikzpicture}
\Biggr\rangle
\end{align}
which energetically penalises configurations of labels around vertices which do not fuse to the identity object.

To define the plaquette operator we first choose an orientation of the plaquette (although the action of $B_p$ is independent of the choice taken). In the following we choose an anti-clockwise convention and define $\{e^{+(-)}\}\in p$ as the set of edges with direction parallel (anti-parallel) to the choice of orientation. We may then write the plaquette operator for $n\in E$ as follows
\begin{align}
&B_{p}^{n}=\Big(\prod_{v\in p}A_{v}\Big)\prod_{e^{+}\in p}\Sigma^{n}_{e}\prod_{e^{-}\in p}\Sigma^{-n}_{e}
\end{align}
where $\Sigma^{n}_{e}$ acts on the label $l$ of edge $e$ such that $\Sigma^{n}_{e}:l\mapsto l+n$. The operators $\Sigma^{n}_{e}$ commute for all edges and $\Sigma^{n}_{e}\Sigma^{m}_{e}=\Sigma^{n+m}_{e}$. The operator $B_p$ in the Hamiltonian is then defined as
\begin{equation}
B_p=\frac{1}{|E|}\sum_{n\in E}B^{n}_{p}.
\end{equation}
As such an operator symmetrises over all group elements the action on basis states is independent of orientation convention for the plaquette.

As the model based on $\mathcal{M}(\mathcal{E})$ does not have any strict dependency on the trivalent lattice we may equally well resolve the trivalent vertices and define the model on a cubic lattice without changing the dynamics of the model. Under such a transformation the vertex operator becomes
\begin{align}
A_v\Biggr|
\begin{tikzpicture}[baseline={([yshift=-25pt]current bounding box.north)}]
\draw[middlearrow={latex reversed}] (0,0) to (0,0.5);
\draw[middlearrow={latex reversed}] (0,0) to (0,-0.5);
\draw[middlearrow={latex reversed}] (0,0) to (-0.5,0);
\draw[middlearrow={latex reversed}] (0,0) to (0.5,0);
\draw[middlearrow={latex reversed}] (0,0) to (0.25,0.25);
\draw[middlearrow={latex reversed}] (0,0) to (-0.25,-0.25);
\node[font=\fontsize{7}{0}\selectfont] at (0,0.65) {$a$};
\node[font=\fontsize{7}{0}\selectfont] at (0,-0.65) {$b$};
\node[font=\fontsize{7}{0}\selectfont] at (-0.65,0) {$c$};
\node[font=\fontsize{7}{0}\selectfont] at (0.65,0) {$d$};
\node[font=\fontsize{7}{0}\selectfont] at (0.35,0.35) {$e$};
\node[font=\fontsize{7}{0}\selectfont] at (-0.35,-0.35) {$f$};
\end{tikzpicture}
\Biggr\rangle&=\delta_{a+b+c+d+e+f,0}\Biggr|
\begin{tikzpicture}[baseline={([yshift=-25pt]current bounding box.north)}]
\draw[middlearrow={latex reversed}] (0,0) to (0,0.5);
\draw[middlearrow={latex reversed}] (0,0) to (0,-0.5);
\draw[middlearrow={latex reversed}] (0,0) to (-0.5,0);
\draw[middlearrow={latex reversed}] (0,0) to (0.5,0);
\draw[middlearrow={latex reversed}] (0,0) to (0.25,0.25);
\draw[middlearrow={latex reversed}] (0,0) to (-0.25,-0.25);
\node[font=\fontsize{7}{0}\selectfont] at (0,0.65) {$a$};
\node[font=\fontsize{7}{0}\selectfont] at (0,-0.65) {$b$};
\node[font=\fontsize{7}{0}\selectfont] at (-0.65,0) {$c$};
\node[font=\fontsize{7}{0}\selectfont] at (0.65,0) {$d$};
\node[font=\fontsize{7}{0}\selectfont] at (0.35,0.35) {$e$};
\node[font=\fontsize{7}{0}\selectfont] at (-0.35,-0.35) {$f$};
\end{tikzpicture}
\Biggr\rangle
\end{align}
while the plaquette operator takes the same form with the trivalent vertex operators replaced with the 6-valent counterpart.

\subsection{Yetter Model for $\mathcal{E}$ on Cubic Lattice}

As mentioned previously in the text the Yetter model can be equally be defined on any cellular decomposition of a 3-manifold. In this section we will outline the model with crossed module of the form of $\mathcal{E}$ on the cubic lattice and show by considering the {\em dual} of the model that such a model is equivalent to the Walker-Wang model of $\mathcal{M}(\mathcal{E})$ on the cubic lattice.

We begin by defining the Yetter model $\mathcal{E}$ on the cubic lattice following the general procedure outlined in section \ref{Hami}. The first step is define an orientation to each square face of the lattice in analogy to the orientation of edges which is inherited from the vertex ordering. Following the previous definitions we choose to orient faces from the lowest ordered vertex on each face which we call the basepoint. We then assign the orientation relative to the two adjacent vertices to the basepoint such that the orientation points to the lowest ordered vertex adjacent to the basepoint. This is demonstrated in the left hand side of equation \eqref{sqorientation} where the face carries the group element $e\in E$ and vertex $a$ is the basepoint and the orientation is given by the relation $a<i<j$. Reversing the orientation of the face replaces the face label with its inverse as shown in the right hand side of equation \eqref{sqorientation}.

\begin{equation}
\begin{tikzpicture}[baseline={([yshift=-54pt]current bounding box.north)}]
\def\q{1.8}
\def\o{0.1}
\draw[] (0,0) to (\q,\q);
\draw[] (\q,\q) to (0,2*\q);
\draw[] (0,2*\q) to (-1*\q,\q);
\draw[] (-1*\q,\q) to (0,0);
\node[font=\fontsize{7}{0}\selectfont] at (0,-0.1) {$a$};
\node[font=\fontsize{7}{0}\selectfont] at (\q+\o,\q) {$i$};
\node[font=\fontsize{7}{0}\selectfont] at (-1*\q-\o,\q) {$j$};
\draw[decoration={markings, mark=at position 1 with {\arrow{latex}}},
postaction={decorate}](0,\q) circle (\q/2);
\node[font=\fontsize{7}{0}\selectfont] at (0,\q) {$e$};
\end{tikzpicture}
=
\begin{tikzpicture}[baseline={([yshift=-54pt]current bounding box.north)}]
\def\q{1.8}
\def\o{0.1}
\draw[] (0,0) to (\q,\q);
\draw[] (\q,\q) to (0,2*\q);
\draw[] (0,2*\q) to (-1*\q,\q);
\draw[] (-1*\q,\q) to (0,0);
\node[font=\fontsize{7}{0}\selectfont] at (0,-0.1) {$a$};
\node[font=\fontsize{7}{0}\selectfont] at (\q+\o,\q) {$i$};
\node[font=\fontsize{7}{0}\selectfont] at (-1*\q-\o,\q) {$j$};
\draw[decoration={markings, mark=at position 1 with {\arrow{latex reversed}}},
postaction={decorate}](0,\q) circle (\q/2);
\node[font=\fontsize{7}{0}\selectfont] at (0,\q) {$e^{-1}$};
\end{tikzpicture}
\label{sqorientation}
\end{equation}

Using the above conventions for the sign of face labels we can now define the 2-flatness condition of a cubic cell. As $\mathcal{E}$ is Abelian and only assigns the identity element to edges, the computation of the 2-holonomy is much simpler than in the general setting. In order to calculate the 2-holonomy of a cubic cell we fix a convention of defining the orientation of faces from either inside or outside of the cubic cell, in the following we choose outside the cell (the flatness condition is independent of such a choice). We then compose the group elements on faces of the cubic with the convention that if the orientation is clockwise we compose the element of the face and if the orientation is anticlockwise we compose the inverse of the face label. We notate this process by introducing the variable $\epsilon\in\{\pm 1\}$ where $\epsilon_{f}=+1(-1)$ if the face $f$ has clockwise (anti-clockwise) orientation such that the 2-holonomy $H_{2}$ on the cube can be written as
\begin{equation}
H_{2}=\sum_{f\in cube}e_{f}^{\epsilon_f}
\end{equation}
and the 2-flatness condition becomes
\begin{equation}
\sum_{f\in cube}e_{f}^{\epsilon_f}=\mathbf{1}_E
\end{equation}

We now define the 2-gauge transformation on the cubic lattice. We may neglect the 1-gauge transform as the 1-gauge group is trivial for the crossed module $\mathcal{E}$. The 2-gauge transformation acts on the four faces adjacent to an edge. We notate the 2-gauge transformation as $A^{h}_{ij}$ on the edge $ij$ where $h\in E$ is the gauge parameter. The gauge transformation has the action of multiplying the faces adjacent to the edge by either $h$ or $h^{-1}$ depending on whether the direction of the edge is parallel or anti-parallel to the orientation of the adjacent edges. An example is shown in equation \eqref{2gaugeaction}.

\begin{align}
A^{h}_{ij}:
\begin{tikzpicture}[baseline={([yshift=-85pt]current bounding box.north)}]
\def\a{1.35}
\draw[middlearrow={latex}] (0,0) to (2*\a,1*\a);
\draw[] (0,0) to (0,1.5*\a);
\draw[] (2*\a,1*\a) to (2*\a,2.5*\a);
\draw[] (0,1.5*\a) to (2*\a,2.5*\a);
\draw[] (0,0) to (2*\a,0);
\draw[] (2*\a,1*\a) to (4*\a,1*\a);
\draw[] (2*\a,0) to (4*\a,1*\a);
\draw[] (0,0) to (0,-1.5*\a);
\draw[dashed] (2*\a,1*\a) to (2*\a,0);
\draw[] (2*\a,0) to (2*\a,-0.5*\a);
\draw[] (0,-1.5*\a) to (2*\a,-0.5*\a);
\draw[] (0,0) to (-2*\a,0);
\draw[dashed] (2*\a,1*\a) to (0,1*\a);
\draw[] (0,1*\a) to (-2*\a,0);
\node[font=\fontsize{7}{0}\selectfont] at (0-0.2,0-0.2) {$i$};
\node[font=\fontsize{7}{0}\selectfont] at (2*\a+0.2,1*\a+0.2) {$j$};
\draw[decoration={markings, mark=at position 1 with {\arrow{latex reversed}}},
postaction={decorate}](1*\a,1.35*\a) circle (0.3*\a);
\node[font=\fontsize{7}{0}\selectfont] at (1*\a,1.35*\a) {$e_1$};
\draw[decoration={markings, mark=at position 1 with {\arrow{latex reversed}}},
postaction={decorate}](1*\a,-0.5*\a) circle (0.3*\a);
\node[font=\fontsize{7}{0}\selectfont] at (1*\a,-0.5*\a) {$e_2$};
\draw[decoration={markings, mark=at position 1 with {\arrow{latex reversed}}},
postaction={decorate}](-0.5*\a,0.35*\a) circle (0.3*\a);
\node[font=\fontsize{7}{0}\selectfont] at (-0.5*\a,0.35*\a) {$e_3$};
\draw[decoration={markings, mark=at position 1 with {\arrow{latex reversed}}},
postaction={decorate}](2.5*\a,0.65*\a) circle (0.3*\a);
\node[font=\fontsize{7}{0}\selectfont] at (2.5*\a,0.65*\a) {$e_4$};
\end{tikzpicture}
\mapsto\no
\begin{tikzpicture}[baseline={([yshift=-85pt]current bounding box.north)}]
\def\a{1.35}
\draw[middlearrow={latex}] (0,0) to (2*\a,1*\a);
\draw[] (0,0) to (0,1.5*\a);
\draw[] (2*\a,1*\a) to (2*\a,2.5*\a);
\draw[] (0,1.5*\a) to (2*\a,2.5*\a);
\draw[] (0,0) to (2*\a,0);
\draw[] (2*\a,1*\a) to (4*\a,1*\a);
\draw[] (2*\a,0) to (4*\a,1*\a);
\draw[] (0,0) to (0,-1.5*\a);
\draw[dashed] (2*\a,1*\a) to (2*\a,0);
\draw[] (2*\a,0) to (2*\a,-0.5*\a);
\draw[] (0,-1.5*\a) to (2*\a,-0.5*\a);
\draw[] (0,0) to (-2*\a,0);
\draw[dashed] (2*\a,1*\a) to (0,1*\a);
\draw[] (0,1*\a) to (-2*\a,0);
\node[font=\fontsize{7}{0}\selectfont] at (0-0.2,0-0.2) {$i$};
\node[font=\fontsize{7}{0}\selectfont] at (2*\a+0.2,1*\a+0.2) {$j$};
\draw[decoration={markings, mark=at position 1 with {\arrow{latex reversed}}},
postaction={decorate}](1*\a,1.35*\a) circle (0.3*\a);
\node[font=\fontsize{7}{0}\selectfont] at (1*\a,1.35*\a) {$e_1-h$};
\draw[decoration={markings, mark=at position 1 with {\arrow{latex reversed}}},
postaction={decorate}](1*\a,-0.5*\a) circle (0.3*\a);
\node[font=\fontsize{7}{0}\selectfont] at (1*\a,-0.5*\a) {$e_2+h$};
\draw[decoration={markings, mark=at position 1 with {\arrow{latex reversed}}},
postaction={decorate}](-0.5*\a,0.35*\a) circle (0.3*\a);
\node[font=\fontsize{7}{0}\selectfont] at (-0.5*\a,0.35*\a) {$e_3-h$};
\draw[decoration={markings, mark=at position 1 with {\arrow{latex reversed}}},
postaction={decorate}](2.5*\a,0.65*\a) circle (0.3*\a);
\node[font=\fontsize{7}{0}\selectfont] at (2.5*\a,0.65*\a) {$e_4-h$};
\end{tikzpicture}
\label{2gaugeaction}
\end{align}

\subsubsection{Model on the Dual Lattice}
\begin{figure}
	\begin{tikzpicture}
	\draw[] (0,0) to (2,1);
	\draw[] (0,0) to (0,1.5);
	\draw[] (2,1) to (2,2.5);
	\draw[] (0,1.5) to (2,2.5);
	\draw[] (0,0) to (2,0);
	\draw[] (2,1) to (4,1);
	\draw[] (2,0) to (4,1);
	\draw[] (0,0) to (0,-1.5);
	\draw[dashed] (2,1) to (2,0);
	\draw[] (2,0) to (2,-0.5);
	\draw[] (0,-1.5) to (2,-0.5);
	\draw[] (0,0) to (-2,0);
	\draw[dashed] (2,1) to (0,1);
	\draw[] (0,1) to (-2,0);
	\draw[thick,blue] (2.2,1.45) to (0.1,1.45); 
	\draw[thick,blue] (-0.1,1.45) to (-0.2,1.45); 
	\draw[thick,blue] (-0.2,1.45) to (-0.2,0.1); 
	\draw[thick,blue] (-0.2,-0.1) to (-0.2,-0.45);
	\draw[thick,blue] (-0.2,-0.45) to (-0.1,-0.45); 
	\draw[thick,blue] (0.1,-0.45) to (2.2,-0.45);
	\draw[thick,blue] (2.2,-0.45) to (2.2,0.05); 
	\draw[thick,blue] (2.2,0.15) to (2.2,1.45);
	\end{tikzpicture}
	\qquad\qquad
	\begin{tikzpicture}[baseline={([yshift=-110pt]current bounding box.north)}]
	\coordinate (A1) at (0, 0);
	\coordinate (A2) at (0, 2);
	\coordinate (A3) at (2, 2);
	\coordinate (A4) at (2, 0);
	
	\coordinate (B1) at (2.05, 1.5);
	\coordinate (B2) at (2.05, 3.5);
	\coordinate (B3) at (4.05, 3.5);
	\coordinate (B4) at (4.05, 1.5);
	
	\draw[] (A1) -- (A2);
	\draw[] (A2) -- (A3);
	\draw[] (A3) -- (A4);
	\draw[] (A4) -- (A1);
	
	\draw[dashed] (A1) -- (B1);
	\draw[dashed] (B1) -- (B2);
	\draw[] (A2) -- (B2);
	\draw[] (B2) -- (B3);
	\draw[] (A3) -- (B3);
	\draw[] (A4) -- (B4);
	\draw[] (B4) -- (B3);
	\draw[dashed] (B1) -- (B4);
	\def\a{0.1};
	\draw[thick,blue] (3.025+\a,2.5) to (1+\a,1);
	\draw[thick,blue] (2.025+\a,2.75) to (2.025+\a,0.75);
	\draw[thick,blue] (1.025+\a,1.75) to (3.025+\a,1.75);
	
	\end{tikzpicture}
	\caption{Examples of the dual of a cubic lattice. The edges of the original lattice are black and the dual edges blue.}
	\label{dual}
\end{figure}
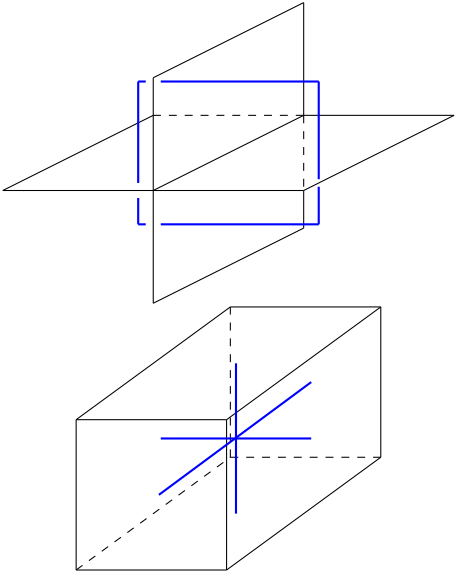

After defining the Yetter model for crossed module $\mathcal{E}$ on the cubic lattice, we will now define the model on the dual cubulation. We define dualisation by a map which takes the n-cells of a cellular decomposition of a d-manifold to the (d-n)-cells of the dual cellulation. We will make the assumption that we are working with a cubulation such that the dual cell decomposition is also a cubulation, such a restriction is for ease of presentation and the arguments follow straightforwardly outside of such a restriction. In this case the cubes (3-cells) are taken to vertices (0-cells) of the new cellulation, square faces (2-cells) are taken to edges (1-cells) and edges (1-cells) are taken to faces (2-cells). In this way we can canonically map the Yetter model with degrees of freedom on faces to a dual lattice where the face labels are now on edges. Examples are shown in figure \ref{dual} where black edges are of the original lattice and blue are dual.

Utilising the duality map discussed previously the direction of dual edges are inherited from the orientation of faces. The direction is defined by the right hand rule, such that if the finger of your right hand point in the direction of the orientation arrow the thumb gives the direction of the dual edge.

\begin{equation}
\begin{tikzpicture}
\def\a{2};
\def\pii{3.14}
\def\ang{\pii/4};
\def\cc{ cos(\ang) };
\def\ss{ sin(\ang) };
\coordinate (A) at ({1*\a*\cc},{1*\a*\ss});
\coordinate (B) at ({-1*\a*\ss},{\a*\cc});
\coordinate (C) at ({1*\a*\ss},{-1*\a*\cc});
\coordinate (D) at ({-1*\a*\cc},{-1*\a*\ss});
\draw[] (A) -- (B);
\draw[] (B) -- (D);
\draw[] (D) -- (C);
\draw[] (C) -- (A);
\draw[thick,middlearrow={latex}] (0,0) -- (\a,\a);
\draw[thick,dotted] (0,0) -- (-1*\a,-1*\a);
\draw [thick,domain=90:360,middlearrow={latex}] plot ({cos(\x)}, {sin(\x)});
\end{tikzpicture}
\end{equation}

Using the above convention to define the directed dual lattice, the Hamiltonian form of the Yetter model for the crossed module $\mathcal{E}$ is vastly simplified. Using the definition of the 2-flatness condition discussed in the previous section, on the dual lattice this constraint becomes the condition
\begin{equation}
\prod_{\tilde{e}\in *(\tilde{v})}g^{\epsilon_{\tilde{e}}}_{\tilde{e}}=\mathbf{1}_E
\end{equation}
where $\tilde{e}$ and $\tilde{v}$ are the dual edges and vertices respectively, $*(\tilde{v})$ is the set of dual edges adjacent to $\tilde{v}$ and $\epsilon_{\tilde{e}}\in{\pm 1}$ is $+1$ when $\tilde{v}$ is the target of $\tilde{e}$ and $-1$ when $\tilde{v}$ is the source. Thus we see that the 2-flatness condition on the cubic cells becomes a vertex condition on the dual lattice. The action of $B^{(2)}_c$ on states of the dual lattice is shown below.
\begin{align}
B^{(2)}_c\Biggr|
\begin{tikzpicture}[baseline={([yshift=-25pt]current bounding box.north)}]
\draw[middlearrow={latex reversed}] (0,0) to (0,0.5);
\draw[middlearrow={latex reversed}] (0,0) to (0,-0.5);
\draw[middlearrow={latex reversed}] (0,0) to (-0.5,0);
\draw[middlearrow={latex reversed}] (0,0) to (0.5,0);
\draw[middlearrow={latex reversed}] (0,0) to (0.25,0.25);
\draw[middlearrow={latex reversed}] (0,0) to (-0.25,-0.25);
\node[font=\fontsize{7}{0}\selectfont] at (0,0.65) {$a$};
\node[font=\fontsize{7}{0}\selectfont] at (0,-0.65) {$b$};
\node[font=\fontsize{7}{0}\selectfont] at (-0.65,0) {$c$};
\node[font=\fontsize{7}{0}\selectfont] at (0.65,0) {$d$};
\node[font=\fontsize{7}{0}\selectfont] at (0.35,0.35) {$e$};
\node[font=\fontsize{7}{0}\selectfont] at (-0.35,-0.35) {$f$};
\end{tikzpicture}
\Biggr\rangle&=\delta_{a+b+c+d+e+f,0}\Biggr|
\begin{tikzpicture}[baseline={([yshift=-25pt]current bounding box.north)}]
\draw[middlearrow={latex reversed}] (0,0) to (0,0.5);
\draw[middlearrow={latex reversed}] (0,0) to (0,-0.5);
\draw[middlearrow={latex reversed}] (0,0) to (-0.5,0);
\draw[middlearrow={latex reversed}] (0,0) to (0.5,0);
\draw[middlearrow={latex reversed}] (0,0) to (0.25,0.25);
\draw[middlearrow={latex reversed}] (0,0) to (-0.25,-0.25);
\node[font=\fontsize{7}{0}\selectfont] at (0,0.65) {$a$};
\node[font=\fontsize{7}{0}\selectfont] at (0,-0.65) {$b$};
\node[font=\fontsize{7}{0}\selectfont] at (-0.65,0) {$c$};
\node[font=\fontsize{7}{0}\selectfont] at (0.65,0) {$d$};
\node[font=\fontsize{7}{0}\selectfont] at (0.35,0.35) {$e$};
\node[font=\fontsize{7}{0}\selectfont] at (-0.35,-0.35) {$f$};
\end{tikzpicture}
\Biggr\rangle
\end{align}
We note $B^{(2)}_c$ now has the same action as the vertex operator in the Walker-Wang model of $\mathcal{M(\mathcal{E})}$.

We now consider the edge gauge transformation. On the dual lattice this operator acts on the four edges bounding a plaquette on the dual lattice $\tilde{p}$. As with the plaquette operator $B_p$ in the Walker-Wang model we define the operator $A_{\tilde{p}}^h$ as the gauge transformation on the dual plaquette $\tilde{p}$ by taking an anti-clockwise orientation around the plaquette and define $\tilde{e^{+(-)}}$ by whether the dual edge $\tilde{e}$ is parallel (anti-parallel) to the orientation convention for the plaquette. We then define $A^{h}_{\tilde{p}}$ as
\begin{equation}
A_{\tilde{p}}^{h}=\prod_{\tilde{e}^{+}\in \tilde{p}}\Sigma^{h}_{\tilde{e}}\prod_{\tilde{e}^{-}\in \tilde{p}}\Sigma^{-h}_{\tilde{e}}
\end{equation}
where $\Sigma^{h}_{e}$ is defined as previously and
\begin{equation}
A_{\tilde{p}}=\frac{1}{|E|}\sum_{h\in E}A_{\tilde{p}}^{h}.
\end{equation}
\subsection{Comparison of Models}

Using the discussion outlined in the previous sections we now compare the Yetter model with input $\mathcal{E}$ and the Walker-Wang model with input $\mathcal{M}(\mathcal{E})$. Both models are defined on a cubic lattice $\Gamma$ with a local Hilbert space defined by $\mathcal{H}=\otimes_{e\in\Gamma}\mathbf{C}^{|E|}$ with edge labels indexed by the group $E$. The Hamiltonian for the Yetter and Walker-Wang models can respectively be written as follows.
\[\begin{array}{l}{\displaystyle
H_{Yetter}(\mathcal{E})=-\sum_{v\in\Gamma}A_v-\sum_{p\in\Gamma}(\frac{1}{|E|}\sum_{h\in E}\prod_{e^{+}\in p}\Sigma^{h}_{e}\prod_{{e}^{-}\in p}\Sigma^{-h}_{e})}\\\\{\displaystyle
H_{WW}(\mathcal{M}(\mathcal{E}))=-\sum_{v\in\Gamma}A_v}\\\\{\displaystyle\sum_{p\in\Gamma}(\frac{1}{|E|}\sum_{h\in E}\prod_{e^{+}\in p}\Sigma^{h}_{e}\prod_{{e}^{-}\in p}\Sigma^{-h}_{e})(\prod_{v\in p}A_v)}
\end{array}\]
Where $A_v=B^{(2)}_c$ is defined on basis states as previous but we reproduce for convenience
\begin{align}
A_v\Biggr|
\begin{tikzpicture}[baseline={([yshift=-25pt]current bounding box.north)}]
\draw[middlearrow={latex reversed}] (0,0) to (0,0.5);
\draw[middlearrow={latex reversed}] (0,0) to (0,-0.5);
\draw[middlearrow={latex reversed}] (0,0) to (-0.5,0);
\draw[middlearrow={latex reversed}] (0,0) to (0.5,0);
\draw[middlearrow={latex reversed}] (0,0) to (0.25,0.25);
\draw[middlearrow={latex reversed}] (0,0) to (-0.25,-0.25);
\node[font=\fontsize{7}{0}\selectfont] at (0,0.65) {$a$};
\node[font=\fontsize{7}{0}\selectfont] at (0,-0.65) {$b$};
\node[font=\fontsize{7}{0}\selectfont] at (-0.65,0) {$c$};
\node[font=\fontsize{7}{0}\selectfont] at (0.65,0) {$d$};
\node[font=\fontsize{7}{0}\selectfont] at (0.35,0.35) {$e$};
\node[font=\fontsize{7}{0}\selectfont] at (-0.35,-0.35) {$f$};
\end{tikzpicture}
\Biggr\rangle&=\delta_{a+b+c+d+e+f,0}\Biggr|
\begin{tikzpicture}[baseline={([yshift=-25pt]current bounding box.north)}]
\draw[middlearrow={latex reversed}] (0,0) to (0,0.5);
\draw[middlearrow={latex reversed}] (0,0) to (0,-0.5);
\draw[middlearrow={latex reversed}] (0,0) to (-0.5,0);
\draw[middlearrow={latex reversed}] (0,0) to (0.5,0);
\draw[middlearrow={latex reversed}] (0,0) to (0.25,0.25);
\draw[middlearrow={latex reversed}] (0,0) to (-0.25,-0.25);
\node[font=\fontsize{7}{0}\selectfont] at (0,0.65) {$a$};
\node[font=\fontsize{7}{0}\selectfont] at (0,-0.65) {$b$};
\node[font=\fontsize{7}{0}\selectfont] at (-0.65,0) {$c$};
\node[font=\fontsize{7}{0}\selectfont] at (0.65,0) {$d$};
\node[font=\fontsize{7}{0}\selectfont] at (0.35,0.35) {$e$};
\node[font=\fontsize{7}{0}\selectfont] at (-0.35,-0.35) {$f$};
\end{tikzpicture}
\Biggr\rangle.
\end{align}

Comparing the two equations above the only difference is in the definition of the second term which acts on plaquettes of the lattice. This difference is actually immaterial as the only distinguishing feature of the term $(\prod_{v\in p}A_v)$ is to increase the energy penalty for colour configurations which do not satisfy the vertex constraint to twice the energy cost of creating plaquette defects. From such a point of view the two Hamiltonians have the same ground-state configurations and the excitations will have the same measurable properties such as braid statistics but the energy cost will be increased for the creation of vertex violations in the Walker-Wang model in comparison to the energy cost in the Yetter model.

\section{Discussion and Outlook}

Here we comment on our main results and discuss the open questions naturally raised by our construction.

The main result of our manuscript is section \ref{Hami}. In this section we outline a large class of exactly solvable Hamiltonian models for topological phases in 3+1D. Such models utilise the conventions of higher lattice gauge theory to define a topological lattice model on a simplicial triangulation of a closed, compact 3-manifold $M$. The algebraic data of the model is defined by a crossed module $\mathcal{G}=(G,E,\partial,\trr)$ (equivalently a 2-group). Each edge  of the triangulation is ``coloured" by a group element $g\in G$ as in topological gauge theory models\cite{kitaev2003fault,moradi2015universal,hu2013twisted,wan2014twisted} while additionally faces of the triangulation are ``coloured" with an element $e\in E$. In a companion paper\cite{BCKMM2} we will further describe the mathematical consistency of our model.

In an additional article\footnote{Twisted Higher Symmetry Topological Phases, Bullivant et al} we will present results further generalising the model in order to include crossed module cohomology utilising the work of Faria Martins and Porter\cite{martins2007yetter}. To extend the model we introduce a 4-cocycle $\omega\in H^4(B\mathcal{G},U(1))$, where $B\mathcal{G}$ is the classifying space of the crossed module $\mathcal{G}$. Such a cocycle adds a $U(1)$ valued phase to the vertex and gauge transformations while the flatness conditions remain unchanged. The value of $\omega$ is determined by considering the gauge transformations as 4-simplices connecting the original lattice colouring to the gauge transformed colouring. $\omega$ is then defined by the 4-cocycle of such a complex. Such a phase generalises the model by allowing for groundstates of the model which are not in an equal superposition of basis states as in the current manuscript.

In section \ref{Yetterrelation} we established the relation between our model and the Yetter Homotopy 2-type TQFT (Yetter TQFT)\cite{yetter1993tqft}. Specifically we showed that the groundstate projector of our model for $M^3$ is given by the Yetter partition function $Z_{Yetter}$ defined on the 4-manifold $M^{3}\times [0,1]$. An intriguing consequence of the proof and the fact that the model can be defined in arbitrary dimension $d>0$ is that this results also holds for arbitrary dimensions $d>0$, but for $d<3$, the Hamiltonian does not contain magnetic operators (there are no blobs in a $d<2$ dimensional boundary). A direct consequence of this is that the groundstate degeneracy ($GSD$) can be obtained by the relation $GSD = Z_{Yetter}(M^d \times S^1)$. We illustrate the formula with several examples in different dimensions for both ordinary and higher gauge theory.

In section \ref{WalkerWang} we described a duality between our model with the crossed module $\mathcal{E}=(\mathbf{1}_{E},E,\partial,\trr)$ and the Walker-Wang model\cite{WalkerWang} with symmetric-braided fusion category $\mathcal{M}(\mathcal{E})$. The duality is established using the results of B\'antay\cite{bantay2005characters} to relate the algebraic data of a crossed module and a braided fusion category. One could expect a further generalisation of such results by noting that the crossed module $M=(G,G,\mathbf{1}_{G},Ad)$ defines a modular tensor category, the quantum double $D(G)$ of the group $G$. This observation is seemingly justified by the fact that both models give rise to a unique groundstate on all 3-manifolds\cite{Simon} although further work is needed to establish such a connection.

Further to the results in this paper, another avenue for exploration would be to classify the excitation spectrum of the model. There are two complimentary approaches to such a classification. The first is to consider local operators of the model\cite{kitaev2003fault}. Using this approach we expect there to be four distinct types of excitations. The four classes of excitations should be point particles and extended line like excitations carrying 1-gauge and 2-gauge charges respectively. Additionally we expect there to be closed loop like excitations and membrane type excitations. The second approach is by considering the quantum numbers associated with the groundstates of our model. In section \ref{gsdsec} we discussed the groundstate degeneracy which is a topological observable of the theory which is independent of local mutations of $M$ which keep the global topology intact, eg. Pachner moves. In general one can consider other topological observables associated with $M$ which come from global transformations of $M$ which keep the global topology invariant. Such global transformations are indexed by the mapping class group (MCG) of $M$. The associated observables should give a full classification of quantum numbers in our model. This approach has already been utilised in\cite{moradi2015universal2,moradi2015universal,wan2014twisted,jiang2014generalized} for 3+1D topological phases using projective representations of $MCG(T^3)=SL(3,\mathbb{Z})$ to understand the quantum numbers for topological gauge theories. $SL(3,\mathbb{Z})$ has two generators given by the $S$ and $T$-matrices. We call eigenstates of the $T$ matrix, $\{\ket{\psi_{j}}\}$ the quasi-particle basis. The eigenvalues of $T$ in such a basis then give the topological spin of the quasi-excitations associated with the groundstate. The exchange statistics of such excitations may be calculated by considering the over-lap of each basis state with the $S$-matrix such that the exchange statistics are given by matrix elements $S_{ij}=\bra{\psi_{j}}S\ket{\psi_{i}}$.

We expect the loop-like excitations of our model to form a representation of the loop-braid group. Note that Yetter's TQFT have been shown to give non-trivial invariants of knotted-surfaces in 3+1D space-time \cite{kauffman2008invariants,faria2009fundamental} and furthermore an embedded $1+1$D TQFT for links in $S^3$ and their cobordisms \cite{Martins2007}.

Another possible generalisation which should be explored in the future is to consider 3-manifolds with boundary. It is known that $BF$ like theories such as the Walker-Wang model with boundaries reproduce chiral anyon theories on 2-dimensional boundaries\cite{Simon}. Such a relation is suggestive that the boundaries of our model could support non-trivial anyon models. 

\bigskip

\noindent {\bf Acknowledgements.}
ZK thanks Kirill Krasnov for a discussion on the Ponzano Regge model. AB would like to thank Jiannis K. Pachos for helpful discussions and Jamie Vicary for introducing the idea of crossed modules.
AB, ZK and PM: \epsrc.\ ZK also thanks the University of Leeds for support under the 
Academic Development Fellowship programme.
MC thanks CNPq for partial financial support via Science Without Borders Programme.
JFM was supported by CMA/FCT/UNL, under the project UID/MAT/00297/2013.
\appendix
\section{\label{a}Algebra of gauge transformations}
In this appendix the proof of the relations (\ref{gr1}) to
(\ref{mixh})
of gauge transformations is presented. 
First we remember that the operators $R_i^g$ and $L_i^g$ are representations of the group $G$, that is,  for all $g,h \in G$ we have $R_i^{gh} = R_i^g R_i^h$ and $L_i^{gh} = L_i^g L_i^h$. Then it is clear that ${\cal{L}}_v^{gh}(i) = {\cal{L}}_v^g(i) {\cal{L}}_v^h(i)$. Also, since $g \trr (h \trr (\cdot)) = gh \trr (\cdot)$, one obtains ${\cal L}_v^{gh}(p) = {\cal L}_v^g(p) {\cal L}_v^h(p)$. Using these properties we deduce that
\[\begin{array}{l}{\displaystyle\prod_{i\in \star{(v)}}{\cal L}^{gh}_v(i)\prod_{p\in \star{'(v)}}{\cal L}^{gh}_v(p) =}\\\\
  {\displaystyle\prod_{i\in \star{(v)}}{\cal L}^g_v(i){\cal L}^{h}_v(i)\prod_{p\in \star{'(v)}}{\cal L}^g_v(p){\cal L}^{h}_v(p)
  =}\\\\{\displaystyle\prod_{i\in \star{(v)}}{\cal L}^g_v(i)\prod_{p\in \star{'(v)}}{\cal L}^g_v(p)
       \prod_{i\in \star{(v)}}{\cal L}^{h}_v(i)\prod_{p\in \star{'(v)}}{\cal L}^{h}_v(p),}\end{array}\]
where in the last equality we used the fact that all the operators in the middle two products commute pairwise, because each one of them acts non-trivially on only a distinct edge or face label. Thus we proved (\ref{gr1}).

Now we prove the identity (\ref{gr2}). First note that $A_i^eA_i^f$ acts as $g_i \mapsto \partial{(e)} \partial{(f)} g_i$ on the edge label of edge $i$, while $A_i^{ef}$ acts as $g_i \mapsto \partial{(ef)} g_i$. Both act trivially on the other edge labels and, since $\partial{(ef)} = (\partial{e}) (\partial{f})$, they have the same action on all edge labels. To prove that they coincide also on face labels, we must consider two cases, (i) $i = v_kv_{k+1}$, and (ii) $i = v_{k+1}v_k$. Also we can suppose that the face $p$ is adjacent to the edge $i$, otherwise both operators act trivially on the edge label of this face.
\begin{enumerate}
	\item[(i)] If $i = v_kv_{k+1}$, $A_i^eA_i^f$ acts as $e_p \mapsto e_p (g_{v_0v_k} \trr f^{-1})(g_{v_0v_k} \trr e^{-1})$ and $A_i^{ef}$ acts as $e_p \mapsto e_p (g_{v_0v_k} \trr (ef)^{-1})$. As $g\trr (\cdot)$ is a homomorphism, we have that the actions coincide.
	\item[(ii)] If $i = v_{k+1}v_k$, $A_i^eA_i^f$ acts as $e_p \mapsto (g_{\overline{v_0v_k}} \trr e)(g_{\overline{v_0v_k}} \trr f) e_p $ and $A_i^{ef}$ acts as $e_p \mapsto (g_{\overline{v_0v_k}} \trr ef)e_p $. As before, the two sides agree. 
\end{enumerate}
So we proved (\ref{gr2}).

In order to verify (\ref{nb1}) we deduce first that $[{\cal L }_v^g(i), {\cal L }_{v'}^h(i)]=0$ and $[{\cal L}_v^g(p),{\cal L}_{v'}^h(p)]=0$, if $v \neq v'$. The first relation holds because both operators can act non-trivially only on the edge $i$ and only if one of the vertices is the source and the other is the target of $i$. However, the operator associated to the source of $i$ is a left multiplication operator and the other associated to the target of $i$ is a right multiplication operator, and these actions are obviously commutative. The second relation follows more easily because at most one of the operators can act non-trivially on a face label. The validity of relation (\ref{nb1}) is now a consequence of the fact that all operators in the definitions of $A^g_v$ and $A^h_{v'}$ commute pairwise.

To prove (\ref{nb2}) we note that $A_i^e$ and $A_{i'}^f$ commute on edge labels because each one of them acts non-trivially on only one edge label and they are distinct. They also commute on a face label if the associated face is not adjacent to both edges, since in this case at least one of them acts trivially on such face label. If the face is adjacent to both edges $i$ and $i'$ and they are oppositely oriented then one is a left action and the other is a right one, so they commute. Assume that $s(i)<t(i)$ and $s(i')<t(i')$. Assume without loss of generality that $s(i')>s(i)$. Then
$A_i^e A_{i'}^f$ acts as $e_p\mapsto e_p (g_{v_0v_{s(i')}}\trr f^{-1}) (g_{v_0v_{s(i)}}\trr e^{-1})$. The action of 
$A_{i'}^f A_i^e$ reads:
\begin{eqnarray*}e_p&\mapsto&e_p(g_{v_0v_{s(i)}}\trr e^{-1})(g_{v_0v_{s(i)}}\partial e\, g_{v_{s(i)}v_{s(i')}}\trr f^{-1})\\
&=&e_p(g_{v_0v_{s(i)}}\trr(e^{-1}(\partial e g_{v_{s(i)}v_{s(i')}}\trr f^{-1})))\\
&=&e_p(g_{v_0v_{s(i)}}\trr(e^{-1}e(g_{v_{s(i)}v_{s(i')}}\trr f^{-1})e^{-1}))\\
&=&e_p(g_{v_0v_{s(i)}}\trr((g_{v_{s(i)}v_{s(i')}}\trr f^{-1})e^{-1}))\\
&=&e_p(g_{v_0v_{s(i')}}\trr f^{-1})(g_{v_0v_{s(i)}}\trr e^{-1})
\end{eqnarray*}
where we used the homomorphism property of $\trr$ in the first and last equation and the second Peiffer condition in the
second. The other case ($s(i)>t(i)$ and $s(i')>t(i')$) is a similar computation. 

Let us consider now the proof of the identities (\ref{mixe}) and (\ref{mixh}). First we note that if $i$ and $v$ are not adjacent to a given face, then $A_i^e$ and $A_v^g$ commute on all edge labels of edges adjacent to this face and on the face label of this face, because in such case at least one of the operators is the identity operator. Therefore, for the rest of the proof we consider a face adjacent to both $i$ and $v$ (if it exists). Note that $A_i^e$ and $A_v^g$ commute on edge labels (on face labels) if $v \neq s(i)$ and $v \neq t(i)$ (if $v \neq v_0$), since in this case $A_v^g$ acts trivially on edge labels (on face labels, 
respectively). Thus we need to verify the identities for the cases: (i) $v = v_0$ and $i=v_0v_1$, (ii) $v=v_0$ and $i=v_0v_{n-1}$, (iii) $v=v_0$, $i \neq v_0v_1$ and $i\neq v_0v_{n-1}$, (iv) $v \neq v_0$ and $v= s(i)$, and (v)  $v \neq v_0$ and $v= t(i)$. Furthermore, it is enough to verify (iii) only on face labels and (iv),(v) only on edge labels. Now we have
\begin{enumerate}
	\item[(i)] $A_i^{g \trr e} A_v^g$ acts as $g_i \mapsto \partial{(g \trr e)} g g_i$ and $e_p \mapsto (g \trr e_p) (g \trr e^{-1})$ and $A_v^g A_i^e$ acts as $g_i \mapsto g\partial{(e)} g_i$ and $e_p \mapsto g \trr (e_p e^{-1})$. They agree since $g \partial{(e)} g_i = g \partial{(e)} g^{-1} g g_i = \partial{(g \trr e)} g g_i$ and $g \trr (e_p e^{-1}) = (g \trr e_p) (g \trr e^{-1})$.
	\item[(ii)] $A_i^{g \trr e} A_v^g$ acts as $g_i \mapsto \partial{(g \trr e)} g g_i$ and 
	$e_p \mapsto (g\trr e)(g\trr e_p)$
	and $A_v^g A_i^e$ acts as $g_i \mapsto g \partial{(e)} g_i$ and $e_p\mapsto g\trr (ee_p)$
	Thus by the previous item the maps agree on edge labels and on face labels by the homomorphisms property of $\trr$. 
	\item[(iii)] For $s(i)<t(i)$ the operator $A_i^{e} A_v^g$ acts as $e_p \mapsto (g \trr e_p) (g g_{v_0s(i)} \trr e^{-1})$ and $A_v^g A_i^e$ acts as $e_p \mapsto g \trr (e_p (g_{v_0s(i)} \trr e^{-1}))$. For $s(i)>t(i)$ the operator $A_i^{e} A_v^g$ acts as $e_p \mapsto (gg_{\overline{v_0v_{s(i)}}^o}\trr e)(g\trr e_p)$ whereas
	$A_v^g A_i^e$ acts as $e_p \mapsto g\trr((g_{\overline{v_0v_{s(i)}}}\trr e)e_p)$
	For both cases equality is clear by the homomorphism property of $\trr$.

	\item[(iv)] $A_i^{g \trr e} A_v^g$ acts as $g_i \mapsto \partial{(g \trr e)} (g g_i)$ and $A_v^g A_i^e$ acts as $g_i \mapsto g \partial{(e)} g_i$. As $\partial{(g \trr e)} (g g_i)= g \partial{e} g^{-1} g g_i = g \partial{e} g_i$, they coincide.
	
	\item[(v)] $A_i^{e} A_v^g$ and $A_v^g A_i^e$ act both as $g_i \mapsto \partial{(e)} g_i g^{-1}$.
\end{enumerate}
This ends the proof of the relations (\ref{gr1}) to (\ref{mixh}).
\section{Transformation properties of 1- and 2-holonomies\label{b}}
In this section we compute the transformation properties of the 1-holonomy of a reference triangle
with labels given by the lhs. of (\ref{reftri}) and a reference tetrahedron with labels given by (\ref{reftet}). 
\allowdisplaybreaks
\begin{eqnarray*}H_1&\xrightarrow{A_a^g}&\partial (g\trr e_{abc})gg_{ab}\,g_{bc}\,g_{ac}^{-1}g^{-1}\\&&=
g(\partial e_{abc}) g_{ab}\,g_{bc}\,g_{ac}^{-1}g^{-1}=gH_1 g^{-1}\\
&\xrightarrow{A_b^g}&(\partial e_{abc}) g_{ab}g^{-1}\,g g_{bc}\,g_{ac}^{-1}=H_1\\
&\xrightarrow{A_c^g}&(\partial e_{abc}) g_{ab}\,g_{bc}g^{-1}\,g g_{ac}^{-1}=H_1\\
&\xrightarrow{A_{ab}^e}&\partial(e_{abc}\,e^{-1})(\partial e)g_{ab}\,g_{bc}\,g_{ac}^{-1}=H_1\\
&\xrightarrow{A_{bc}^e}&\partial(e_{abc}\,g_{ab}\trr e^{-1})g_{ab}\,\partial e\,g_{bc}\,g_{ac}^{-1}\\&&=
\partial e_{abc}\,g_{ab}\,\partial e^{-1}g_{ab}^{-1}\,g_{ab}\,\partial e\,g_{bc}\,g_{ac}^{-1}=H_1\\
&\xrightarrow{A_{ac}^e}&\partial (e\,e_{abc})g_{ab}\,g_{bc}\,g_{ac}^{-1} \partial e^{-1}=\partial e H_1 (\partial e)^{-1}\\
H_2&\xrightarrow{A_a^g}&(g\trr e_{acd})(g\trr e_{abd})(g g_{ab}\trr e_{bcd}^{-1})(g\trr e_{abd})=
g\trr H_2\\
&\xrightarrow{A_b^g}&e_{acd}\,e_{abc}(g_{ab}g^{-1}\trr (g\trr e_{bcd}^{-1}))e_{abd}^{-1}=H_2\\
&\xrightarrow{A_c^g}&H_2\\
&\xrightarrow{A_d^g}&H_2\\
&\xrightarrow{{ A}_{ab}^e}&e_{acd}\,e_{abc} e^{-1}((\partial e g_{ab})\trr e_{bcd}^{-1})e_{abd} e^{-1})^{-1}=\\&&
e_{acd}\,e_{abc} e^{-1}[(\partial e g_{ab})\trr e_{bcd}^{-1}] e\,e_{abd}^{-1}=\\
&&e_{acd}\,e_{abc} e^{-1}(\partial e\trr (g_{ab}\trr e_{bcd}^{-1}))e e_{abd}^{-1}=\\&&
e_{acd}\,e_{abc} e^{-1}e(g_{ab}\trr e_{bcd}^{-1}))e^{-1} e e_{abd}^{-1}=H_2\\
&\xrightarrow{A_{bc}^e}&e_{acd}\,e_{abc}(g_{ab}\trr e^{-1})(g_{ab}\trr e\,e_{bcd}^{-1})e_{abd}^{-1}\\&&=
e_{acd}\,e_{abc}(g_{ab}\trr (e^{-1}e e_{bcd}^{-1})) e_{abd}^{-1}=H_2\\
&\xrightarrow{A_{ac}^e}&e_{acd}e^{-1}e e_{abc}(g_{ab}\trr e_{bcd}^{-1})e_{abd}^{-1}=H_2\\
&\xrightarrow{{ A}_{ad}^e}&e\,e_{acd}\,e_{abc}(g_{ab}\trr e_{bcd}^{-1})\,(ee_{abd})^{-1}=e H_2\,e^{-1}\\
&\xrightarrow{A_{bd}^e}&e_{acd}\,e_{abc}(g_{ab}\trr(e e_{bcd})^{-1})(e_{abd}(g_{ab}\trr e^{-1}))^{-1}=\\&&
e_{acd}\,e_{abc}(g_{ab}\trr(e_{bcd}^{-1}e^{-1})(g_{ab}\trr e)e_{abd}^{-1}=\\
&&e_{acd}\,e_{abc}(g_{ab}\trr(e_{bcd}^{-1}e^{-1}e)e_{abd}^{-1}=H_2\\
&\xrightarrow{{ A}_{cd}^e}&e_{acd}(g_{ac}\trr e^{-1})e_{abc}\,[g_{ab}\trr(e_{bcd}(g_{bc}\trr e^{-1}))^{-1}]e_{abd}^{-1}\\&&=
e_{acd}(g_{ac}\trr e^{-1})e_{abc}\,[g_{ab}\trr (g_{bc}\trr e) e_{bcd}^{-1}]e_{abd}^{-1}=\\&&
e_{acd}(g_{ac}\trr e^{-1})e_{abc}(g_{ab}\,g_{bc}\trr e)(g_{ab}\trr e_{bcd}^{-1})e_{abd}^{-1}=\\&&
e_{acd}\left(\left(g_{ac}(g_{ab}\,g_{bc})^{-1}
\right)\trr e'^{-1}\right)\times\\&&e_{abc}e'(g_{ab}\trr e_{bcd}^{-1}) e_{abd}^{-1}=\\
&&e_{acd}((H_1^{abc})^{-1}\partial e_{abc})\trr e'^{-1}\times\\&&e_{abc}e'(g_{ab}\trr e_{bcd}^{-1}) e_{abd}^{-1}\approx\\&&
e_{acd}\,e_{abc}e'^{-1}e_{abc}^{-1}e_{abc}e'(g_{ab}\trr e_{bcd}^{-1}))e_{abd}^{-1}=H_2
\end{eqnarray*}

In the last equation the substitution $e'=g_{ab}\,g_{bc}\trr e$ has been made and $\approx$ means equality in case 
when $H_1^{abc}=1$. Note that the very last relation shows that the 2-holonomy does not transform covariantly if fake flatness of the
boundary faces is not imposed.

\section{Proof of Theorem \ref{th1}\label{c}}
A three cell of $\Delta$ is a prism based on 
a face of $L_j$.
\begin{center}
	\includegraphics{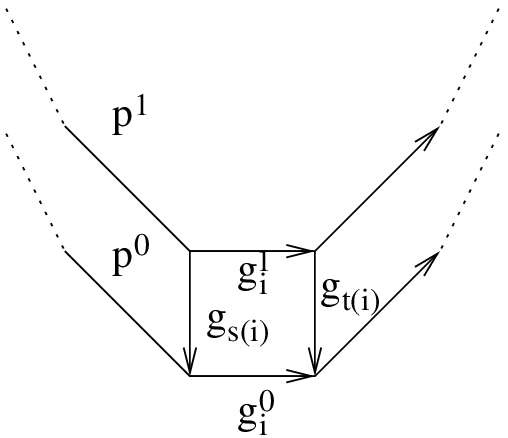}
\end{center}
The figure shows a part of the complex. 
The lattice $\Delta$ consists of $L_{0c}$, $L_{1c}$ with the coloring given 
by $\{g_i^0\}_{i\in L^1_0}$, $\{g_i^1\}_{i\in L^1_1}$, $\{g_v\}_{v\in L^0}$, oriented edges in 
$L_0,L_1$ and vertical edges assumed to be oriented towards $L^0$ 
(downwards in the figure) connecting corresponding vertices of $L_j$, respectively. 

We have a Dirac delta $\delta_{g_p,{\bf 1}}$ in the partition function for each face. 
In particular, for an internal face $p$ (this is always a rectangle connecting corresponding edges of $L_j$) 
the term $\delta_{g_p,{\bf 1}}$ enforces 
$g^1_i=g_{s(i)}g_i^0 g_{t(i)}^{-1}$. Taking all such faces into account, identifying the colouring 
$L_{0c}$ with $|L_{0c}\rangle$ 
we have\footnote{Note, that $|L_{0c}\rangle$ is a basis element, not a generic state
	in the Hilbert space.}
\[ \prod_{p\in \Delta^{(i)2}} \delta_{g_p,{\bf 1}}=\langle L_{1c}|\prod_{v\in L^0} A^{g_v}_v|L_{0c}\rangle\]
and consequently we can write
\begin{align}&\frac{1}{|G|^{|L^0|}}
\prod_{v\in L^0}\sum_{g_v\in G}
\prod_{p\in \Delta^{(i)2}} \delta_{g_p,{\bf 1}}\\=&\bigg\langle L_{1c}\bigg|\prod_{v\in L^0} \frac{1}{|G|}\sum_{g_v\in G}
A^{g_v}_v\bigg|L_{0c}\bigg\rangle=\bigg\langle L_{1c}\bigg|\prod_{v\in L^0}A_v\bigg|L_{0c}\bigg\rangle\label{inte}\end{align}
Now, let us compute the prefactor in the definition (\ref{bzlgt}) for the lattice: 
$|G|^{-\frac{|L^0_0|+|L^0_1|}{2}}=|G|^{-|L_0|}$. Hence the lhs. of the formula above  almost coincides with the
lhs. of (\ref{lgt}), only $\prod_{p\in\Delta^{(b)2}}\delta_{g_p,{\bf 1}}$ is missing. This enforces flatness on the faces 
$p\in L_0^2\cup L_1^2$. It agrees with the action of the operator $\prod_{p\in L^2}B_p$ on $|L_{0c}\rangle$ times that on 
$|L_{1c}\rangle$ (note that 
the lattices $L_0$ and $L_1$ are identified, but the states $|L_{0c}\rangle$ and $|L_{1c}\rangle$ are different). However,
the operators $B_p, p\in L^2$ and $A_v, v\in L^0$ commute for any pair of labels and are also self-adjoint, so we can simply 
insert the factor $\prod_{p\in L^2}B_p$ 
anywhere in the scalar product, using the fact that $B_p^2=B_p, p\in L^2$, thus inserting $B_p$ only once is sufficient.\qed

\section{Proof of Theorem \ref{th2}\label{d}}
Consider that $M^3\times [0,1]$ has the product lattice decomposition $\Delta$.  Let us choose a total order on $\Delta^0$ in the following way. The Hilbert space 
is associated to a dressed $L$, i.e., $L^0$ has a total order. Let the total orders in 
$L_i^0$ agree with that on $L^0$ and let any vertex in $L_1^0$ be smaller than any other in 
$L_0^0$. Then denoting the colour of the internal edge connecting the vertices in $L^0_0$ and $L^0_1$ 
corresponding to $v\in L^0$ 
by $g_v$ and the colour of the internal face connecting edges in $L_0^1$ and $L_1^1$ 
corresponding to $i\in L^1$ by $e_i$,
the following equality is true.
\begin{equation}\begin{array}{l}{\displaystyle 
\prod_{P\in\Delta^{3(i)}}\delta_{H_2(P),{\bf 1}}\prod_{p\in \Delta^{2(i)}} \delta_{H_1(p),{\bf 1}}=}\\\\{\displaystyle\bigg \langle L^{c_1}\bigg|\prod_{v\in L^0} A^{g_v}_v \prod_{i\in L^1}A^{e_i}_i\bigg|L^{c_0}\bigg \rangle}
\end{array}\label{gp}\end{equation} 
\begin{figure}\caption{\label{p2}
		The left figure shows one of the internal faces connecting corresponding boundary edges of $L_j$, 
		the top edge label $g_i^1$ is determined by fake flatness of the rectangle. By our choice of total order on $\Delta^0$, the 
		basepoint of the rectangle is $s(i)$ and the fake flatness constraint reads as the composition (\ref{fifo}). 
		The right figure shows a blob bounded by the "bucket" consisting of the green pentagon of $L_0$ with label $e_p^0$ and the
		rectangles with labels $e_i\in E,i\in bd(p_0)$. The 2-holonomy of the bucket is $\tilde{e}_p^0$ the "lid pentagon" is coloured by
		$e_p^1$. By 2-flatness of the blob $\tilde{e}_p^0=e_p^1$.
	}\includegraphics[width=8cm]{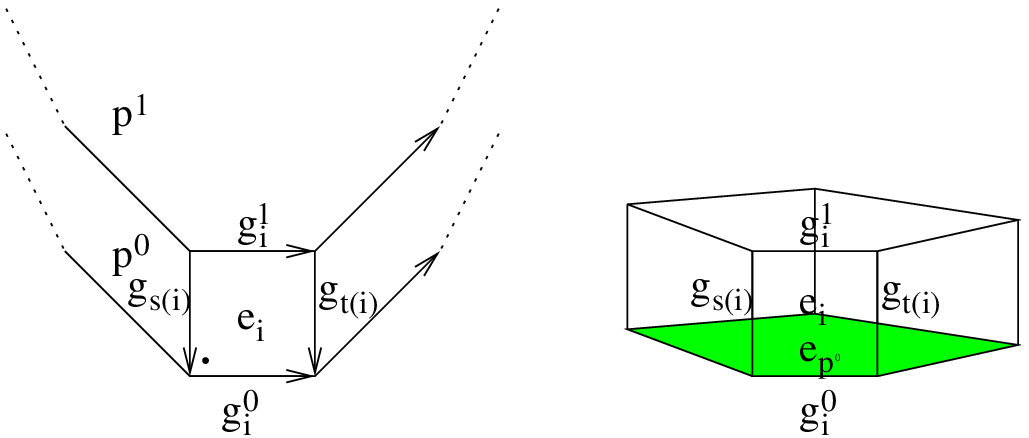}\end{figure}
To justify it, let us consider the rectangle $p$ depicted in Fig.\ref{p2} with boundary edges 
$g_i^0,g_{s(i)},g_{t(i)},g_i^1$. The constraint
$\delta_{H_1(p),{\bf 1}}$ enforces $\partial e_i=g_{s(i)}^{-1}g_i g_{t(i)} (g_i^0)^{-1}$. Equivalently
\begin{equation} g^1_i=g_{s(i)}\partial e_i g_i^0 g_{t(i)}^{-1}\ , \label{fifo}\end{equation}
which is precisely the image of $g_i^0$ under the product of gauge transformations with parameters $g_{s(i)}, g_{t(i)}, e_i$ at vertices
$s(i), t(i)$ and edge $i$, respectively such that the edge transformation acts first. Note that no other generators in the product act 
on the edge color $g_i$. Let us recall that we identify $L_{jc}$ with the vector $|L_{jc}\rangle$ and 
colours on $\Delta^{(i)}$ are identified with parameters of gauge transformation on the Hilbert space.  

Now consider $p^0\in L_{0c}$ 
coloured with $e_p^0$ corresponding to the 2-holonomy based at $v_0(p^0)$ 
and denote the boundary edge set by $L^1_{p^0}$. Assume that $p^0$ is an $n$-gon. 
Denote the disk (depicted as a bucket in Fig.\ref{p2}) by $\tilde{p}^0$, consisting of the rectangles bounded by
the edges $\{g_i^0,g_{s(i)},g_{t(i)},g_i\}, i\in bd(p^0)$ and $p^0$ . We can compute the 2-holonomy 
of this disk based at $v_0(p)$, which we denote by $e^0_{\tilde{p}}$. 
We have to show that this (again via the identification of boundary colourings with basis vectors in the Hilbert space and
internal colours as parameters of gauge transformations) agrees with the action of the product
$\prod_{v\in L^0} A^{g_v}_v \prod_{i\in L^1}A^{e_i}_i$ on $e^0_p$ where $g_v$ is the color of the vertical edge pointing toward
the vertex $v$ and $e_i$ is the color of the vertical rectangle based at edge $i$. Introducing the notation $i$ 
for the $i$-th edge starting
from $v_0(p)$ in the circular order 
the action of the product of edge transformations reads
\begin{align}&e^0_p\mapsto e'^0_p\equiv\nonumber\\&
e_{p_0}(\partial e_{p_0}^{-1}\trr e_n)(g_{v_0,s(n-1)}\trr e^{\pm 1}_{n-1})(g_{v_0,s(n-2)}\trr e^{\pm 1}_{n-2})\dots
\nonumber\\&\dots (g_{v_0,s(2)}\trr e^{\pm 1}_2)e_1^{-1}\ ,\label{gspc}\end{align}
where $e_i^{\pm 1}$ means $e_i$ ($e_i^{-1}$), if $i$ is oriented opposite (according) to the orientation of $bd(p)$, respectively. 
Now we apply the vertex transformations.  
We will show that (i) the rhs. of (\ref{gspc}) is unchanged under a vertex gauge transformation 
$A_v^{g_v}$ with $v\neq v_0$ and (ii) that it changes
as $(.)\mapsto g_{v_0}\trr (.)$ for $v=v_0$. 
This is how the face holonomy $e'^0_p$ should change under $\prod_{v\in L^0} A_v^{g_v}$. Consequently, 
the 2-holonomy $e^0_{\tilde p}$ of the disk $\tilde{p}^0$ is given by the image of $e_p^0$ under gauge transformation.

It is easy to see that $g_{v_0,s(k)}$ changes only if $v=s(k)$ since then 
$g_{s(k)-1,s(k)}\mapsto g_{s(k)-1,s(k)}g_v^{-1}$, which induces $g_{{v_0},s(k)}\mapsto g_{{v_0},s(k)}g_v^{-1}$. 
The label $e_k$ changes also only for $s(k)=v$ as $e_k\mapsto g_v\trr e_k$. Altogether
\[ g_{{v_0},s(k)}\trr e_k\mapsto g_{{v_0},s(k)}g_v^{-1}\trr (g_v\trr e_k)=g_{{v_0},s(k)}\trr e_k\]
so (i) is proved. The action of $A_{v_0}^{g_{v_0}}$ is $e_{p_0}\mapsto g_{v_0}\trr e_{p_0}$ and $e_1\mapsto g_{v_0}\trr e_1$ on the
the two faces with basepoint $v_0$ and all parenthesis in (\ref{gspc}) transform as $(.)\mapsto g_{v_0}\trr (.)$ since 
$g_{{v_0},s(k)}\mapsto g_{v_0}g_{{v_0},s(k)}$. Hence, since $(.)\mapsto g\trr(.)$ is a homomorphism, we showed (ii), so the claim
is proved. Note that the edge labels change as $g^0_i\mapsto\partial e_i g^0_i\mapsto g_{s(i)}\partial e_i g^0_i 
g_{t(i)}^{-1}$ under $\prod_j A_j^{e_j}$ and $\prod_v A_v^{g_v}$, respectively, in accordance with (\ref{fifo}).
Another remark is that the order of the terms on the rhs. of (\ref{gp}) depends on the total order on $\Delta^0$. In 
particular, if we had chosen $v^0_{s(i)}>v^1_{s(i)}$ the fake flatness condition would have read 
$\partial e_i=g_i^1 g_{t(i)}(g_i^0)^{-1}g_{s(i)}^{-1}$ equivalent to the order $A_i^{e_i}A_{s(i)}^{g_{s(i)}}$ of action
of gauge transformations. Once the integration is done, this dependence will of course disappear, equivalently the 
vertex projections defined as averaged gauge transformations mutually commute.

Now, consider the 2-sphere $S_{p_0}$ 
bounded by the disks $\tilde{p}^0$ and $p^1$ coloured by $e^0_{\tilde{p}}$ and $e^1_p$, respectively.
They have the same boundary $bd(p^1)$, and they are oriented oppositely. Hence, 
$\delta_{H_2(S_{p_0}),{\bf 1}}=1$ iff $e^0_{\tilde{p}}=e_{p^1}$. This completes the proof of (\ref{gp}).

Let us now determine the multiplicative factors from the definition 
of the partition vector from (\ref{bzlgt}) for the lattice at hand. We have only boundary vertices 
$\Delta^0=L_0^0\cup L_1^0$ and the edge set is in one-to-one correspondence with $L^0\cup (L_0^1\cup L_1^1)$ such that the first 
factor is for internal edges and the second for external ones, respectively. The multiplicative factor determined from 
(\ref{bzlgt}) turns out to be $|G|^{-|L^0|}|E|^{-|L^1|}$. This means that we can assign a factor of $|G|^{-1}$ to 
each term $\sum_{g_v} A_v^{g_v}, v\in L^0$ and a factor of $|E|^{-1}$ to $\sum_{e_i} A_i^{e_i}, i\in L^1$. So, similarly
to the ordinary gauge theory case, we found
\begin{align*} 
&\frac{1}{|G|^{|L^0|}|E|^{|L^1}}\prod_{v\in L^0}\prod_{i\in L_0^1}\sum_{g_v\in G}\sum_{e_i\in E}\\
&\prod_{P\in\Delta^{3(i)}}\delta_{H_2(P),{\bf 1}}\prod_{p\in \Delta^{2(i)}} \delta_{H_1(p),{\bf 1}}=\\
&\bigg\langle L^{c_1}\bigg|\prod_{v\in L^0} A_v \prod_{i\in L^1}A_i\bigg|L^{c_0}\bigg\rangle
\end{align*} 

The last step is the identification of the 2-flatness constraints of the
blobs in $L_0^3\cup L_1^3$ with the $B_P$ operators. This goes parallel to the ordinary lattice gauge theory case. \qed
\bibliographystyle{apsrev4-1}
\bibliography{Enriched-Kitaev-Model-Notes.bib}

\end{document}